\documentclass[10pt, conference, letterpaper]{ IEEEtran}
\IEEEoverridecommandlockouts

\usepackage{cite}
\usepackage{amsmath,amssymb,amsfonts}
\usepackage{graphicx}
\usepackage{textcomp}
\usepackage{xcolor}
\usepackage{calligra}
\usepackage{algorithm}
\usepackage{algorithmicx}
\usepackage{algpseudocode} 
\usepackage{amsmath} 
\usepackage{subfigure}
\usepackage{float}
\usepackage{etoolbox}
\usepackage{tabu}
\usepackage{booktabs}
\usepackage[table]{xcolor}
\usepackage{lipsum}

\usepackage[T1]{fontenc}
\def\BibTeX{{\rm B\kern-.05em{\sc i\kern-.025em b}\kern-.08em
    T\kern-.1667em\lower.7ex\hbox{E}\kern-.125emX}}
\begin{document}

\title{AirGS: Real-Time 4D Gaussian Streaming for Free-Viewpoint Video Experiences \\

\thanks{This work is supported by the National Key R\&D Program of China (Grant No. 2024YFC3017100) and NSFC No. 62302292. \IEEEauthorrefmark{1}Corresponding author.}
}

\author{\IEEEauthorblockN{Zhe Wang, Jinghang Li and Yifei Zhu\IEEEauthorrefmark{1}}
\IEEEauthorblockA{\textit{Global College, Shanghai Jiao Tong University } \\
Email: 123369423@sjtu.edu.cn, sjtulijinghang@sjtu.edu.cn, yifei.zhu@sjtu.edu.cn}
}


\maketitle

\begin{abstract}

Free-viewpoint video (FVV) enables immersive viewing experiences by allowing users to view scenes from arbitrary perspectives. As a prominent reconstruction technique for FVV generation, 4D Gaussian Splatting (4DGS) models dynamic scenes with time-varying 3D Gaussian ellipsoids and achieves high-quality rendering via fast rasterization. However, existing 4DGS approaches suffer from quality degradation over long sequences and impose substantial bandwidth and storage overhead, limiting their applicability in real-time and wide-scale deployments. Therefore, we present AirGS, a streaming-optimized 4DGS framework that rearchitects the training and delivery pipeline to enable high-quality, low-latency FVV experiences. AirGS converts Gaussian video streams into multi-channel 2D formats and intelligently identifies keyframes to enhance frame reconstruction quality. It further combines temporal coherence with inflation loss to reduce training time and representation size. To support communication-efficient transmission, AirGS models 4DGS delivery as an integer linear programming problem and design a lightweight pruning level selection algorithm to adaptively prune the Gaussian updates to be transmitted, balancing reconstruction quality and bandwidth consumption. Extensive experiments demonstrate that AirGS reduces quality deviation in PSNR by more than 20\% when scene changes,  maintains frame-level PSNR consistently above 30, accelerates training by 6$\times$, reduces per-frame transmission size by nearly 50\% compared to the SOTA 4DGS approaches.

\end{abstract}


\section{Introduction}

Free-viewpoint video (FVV) allows users to explore a dynamic scene from arbitrary viewpoints, offering an immersive and interactive visual experience. To generate the FVV, the straightforward way is to reconstruct the scene into 3D model from captured multi-view 2D image sequences. 
Existing dynamic scene reconstruction methods either explicitly model scene geometry using volumetric or mesh-based primitives \cite{dou2017motion2fusion, newcombe2015dynamicfusion}, or synthesize novel views through image-based interpolation \cite{broxton2020immersive, zitnick2004high, li2021neural}. 
However, these methods often struggle to achieve high reconstruction quality in real-world scenes with complex geometry and rich appearance variations \cite{sun20243dgstream}.
\begin{figure*}[!tbp]
    \centering
    \subfigure[Dynamic scene input]{\includegraphics[height=3.3cm,width=0.21\linewidth]{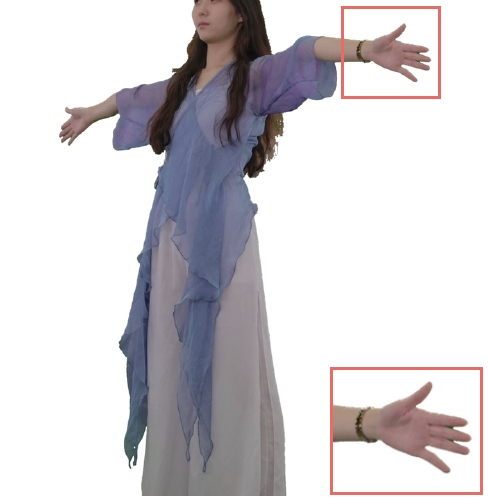}\label{GT_v_d}}
    \subfigure[3DGStream \cite{sun20243dgstream}]{\includegraphics[height=3.3cm,width=0.21\linewidth]{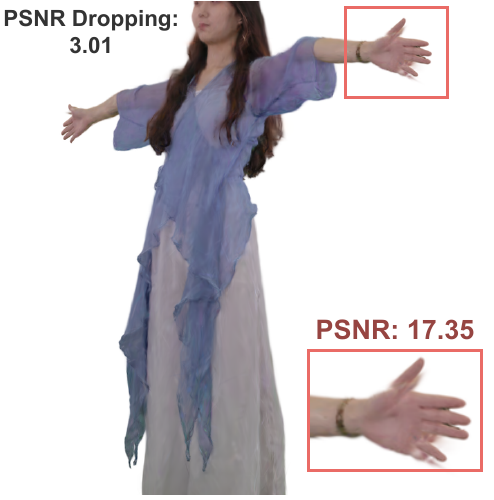}\label{3dgstream_v_d}}
    \subfigure[$V^3$ \cite{wang2024v}]{\includegraphics[height=3.3cm,width=0.21\linewidth]{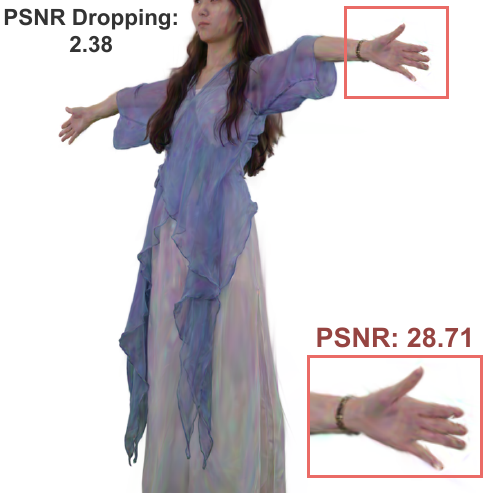}\label{v3_v_d}}
    \subfigure[The proposed method: AirGS]{\includegraphics[height=3.3cm,width=0.21\linewidth]{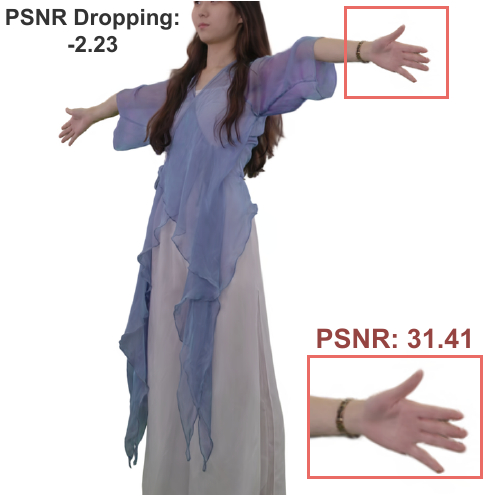}\label{airgs_v_d}}
    \caption{We present AirGS, a novel framework for real-time, high-quality 4DGS-based FVV generation and streaming in practical networks. The PSNR dropping refers to the difference between the reconstruction quality of the first frame and that of the current frame (the 25th frame in this example). The red PSNR value indicates the quality of the region highlighted by the red box.}
    \label{background_part}
\end{figure*}

Recently, 3D Gaussian Splatting (3DGS) \cite{kerbl20233d} has emerged as a highly efficient method for novel view synthesis, offering superior rendering quality and speed. The core idea is to represent a static scene as a set of 3D Gaussian ellipsoids learned via optimization, and to render them efficiently using fast rasterization. Building on this breakthrough, several efforts have extended 3DGS to 4D dynamic scene modeling \cite{wu20244d, yang2023real}. However, these methods typically use a fixed number of Gaussian ellipsoids across all frames, which limits reconstruction quality over long sequences and increases per-frame storage overhead. To address this, 3DGStream \cite{sun20243dgstream} introduces a keyframe-based design, where a reference frame (anchor) is represented with full 3D Gaussians, and subsequent frames capture dynamics via a lightweight MLP that predicts Gaussian updates in the anchor’s space. To reduce the runtime cost associated with frequent MLP queries, $V^3$ \cite{wang2024v} further stores precomputed MLP outputs as multi-channel 2D images, enabling faster decoding with minimal computational overhead.

These advancements demonstrate the potential of 4DGS-based FVV generation. However, streaming 4DGS content in practical network environments remains largely unexplored. Fundamentally, several key challenges persist towards achieving real-time deployment. 
First, while the introduction of keyframe and the auxiliary MLP significantly reduces communication cost, keyframe selection remains an open problem.  Most of existing studies fix the set of Gaussian primitives in the anchor space after initialization. Consequently, simply selecting the first frame as a keyframe often leads to noticeable quality degradation and loss of image details, demonstrated in Fig. \ref{background_part}, since early frames may lack information about later-appearing objects or large motions. 
Second, training a full 3DGS model for each keyframe is computationally expensive. 
Naively increasing the number of keyframes improves reconstruction quality, but significantly increases training time and resource usage. 
Both highlight the need for an efficient and scalable training strategy.
Third, transmitting a full 3DGS for every frame incurs substantial bandwidth and latency overhead, making it difficult to satisfy real-time requirements in practical fluctuating networks. Therefore, a communication-efficient streaming framework is also needed to ensure a smooth and high-quality FVV experience.

\begin{figure}[hbt!]
        \centering
 
        \subfigure[3DGS]{
            \begin{minipage}[b]{0.95\linewidth}
            \includegraphics[width=0.95\linewidth]{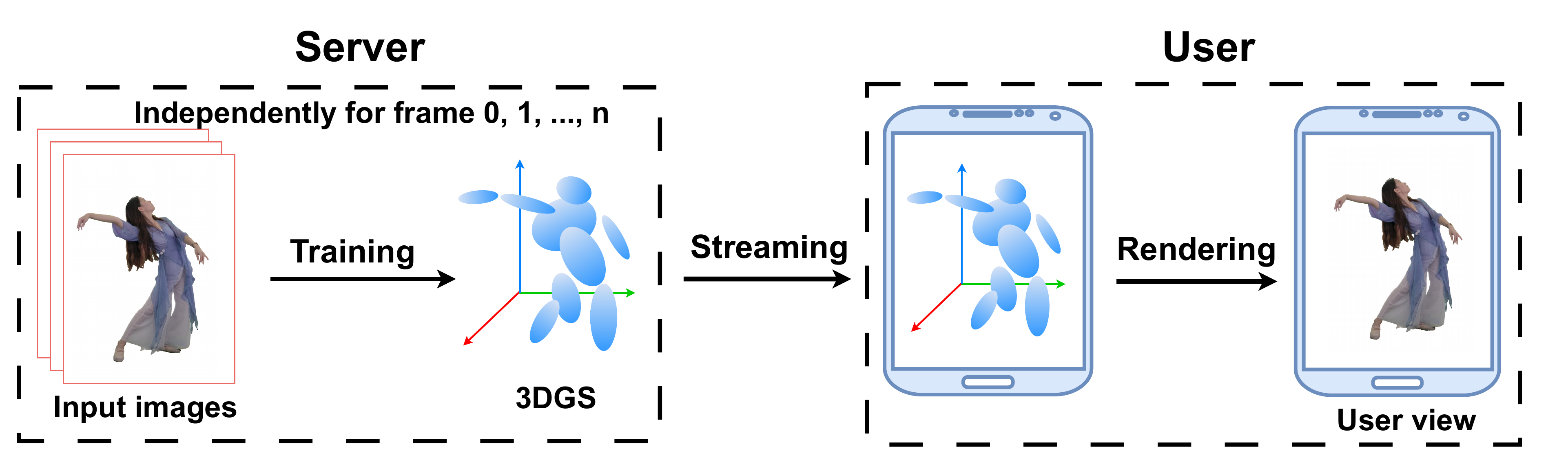}
            \label{subfig:3dgs}
            \end{minipage}
            }
        \subfigure[Typical 4DGS]{
            \begin{minipage}[b]{0.95\linewidth}
            \includegraphics[width=\linewidth]{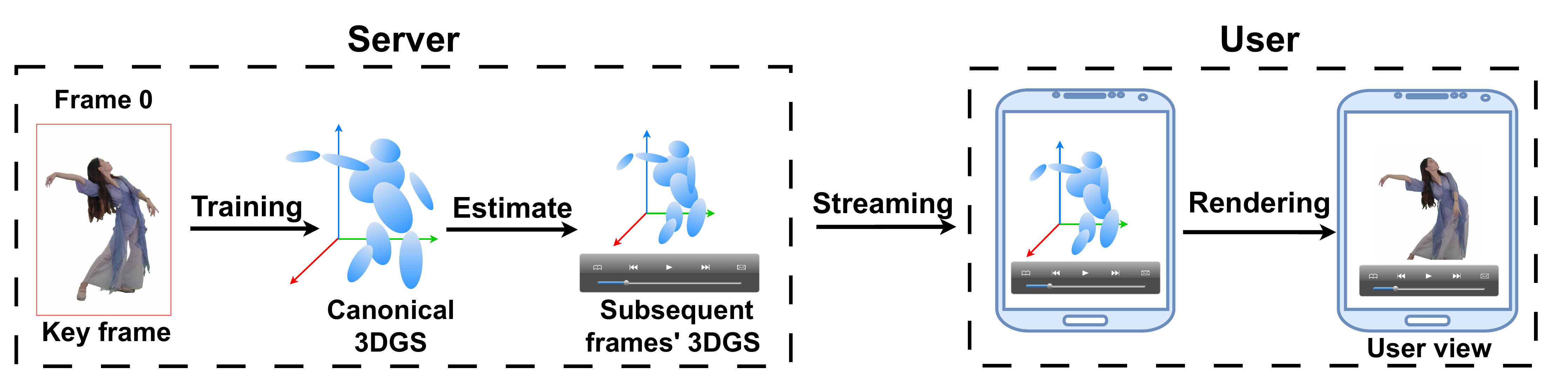}
            \label{subfig:4dgs}
            \end{minipage}
            }
        \subfigure[Streamable 4DGS]{
            \begin{minipage}[b]{0.95\linewidth}
            \includegraphics[width=0.95\linewidth]{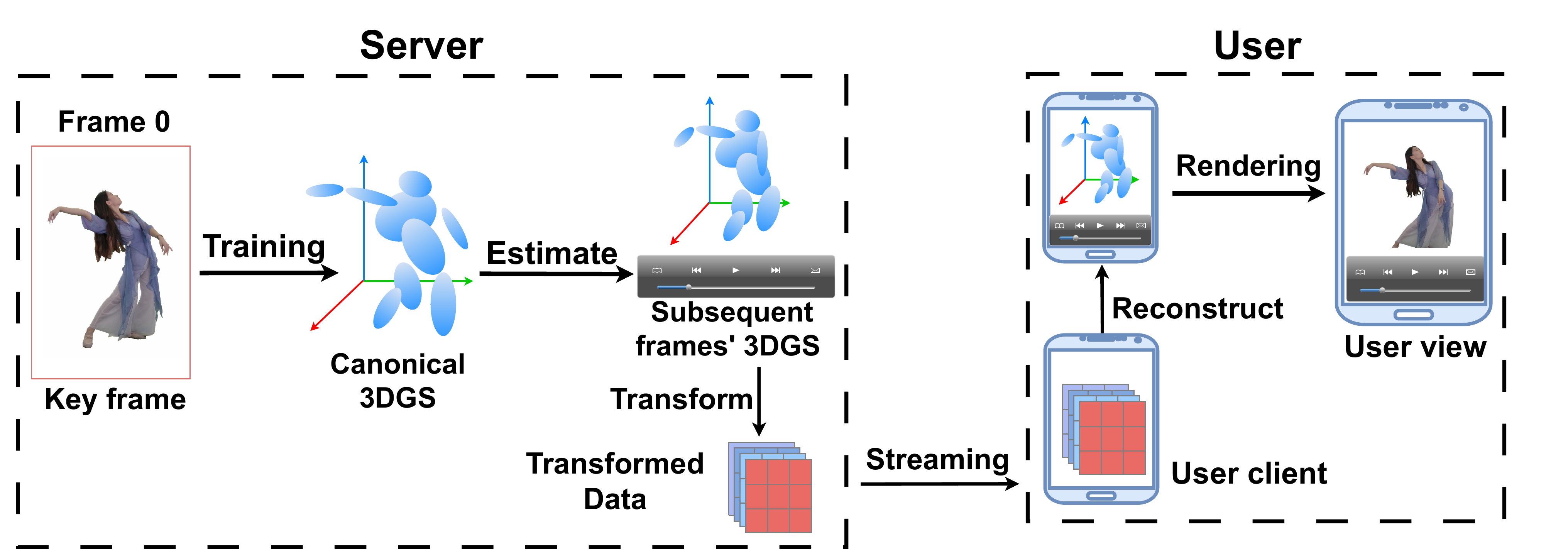}
            \label{subfig:airgs}
            \end{minipage}
            }
        \caption{Illustration of 3DGS, typical 4DGS and streamable 4DGS for FVV generation}
\end{figure}

To address these challenges, we propose AirGS, a Gaussian-based 4D video streaming framework designed for high-quality  reconstruction over long-sequence and smooth rendering under fluctuating network conditions. 
AirGS builds upon the idea of converting the dynamic Gaussian sequence into multi-channel long-sequence 2D Gaussian frames, enabling compatibility with existing video codecs and facilitating efficient streaming.
It systematically redesigns two fundamental components of the 4DGS pipeline: the training strategy and the streaming framework. In training, AirGS intelligently selects frames with quality degradation as keyframes to initiate new canonical spaces.
To reduce retraining overhead, it exploits temporal continuity by incorporating the neighboring frame of the keyframe and introduces a novel inflation loss to accurately reconstruct new objects and large motions while minimizing additional Gaussian primitives and training costs. For streaming, AirGS transmits only Gaussian deltas between consecutive frames, leveraging the inter-frame similarity and sparsity we observe in 4DGS.
It further incorporates a frequency-based importance evaluator and a network-aware Gaussian-pruner to adapt to the fluctuating networks, reducing transmission overhead while preserving rendering fidelity. To our knowledge, AirGS is the first streaming-optimized 4DGS-based free-viewpoint video generation framework. 
In summary, our contributions are as follows:
\begin{itemize}
    \item We propose AirGS, a novel streaming-optimized 4DGS framework that enables high-quality real-time FVV generation and streaming over dynamic networks.
    \item We propose a quality-driven keyframe selection strategy to capture abrupt object appearances and large motions. We further exploit temporal continuity and introduce a Gaussian inflation loss to further reduce training cost and representation size.
    \item We design an adaptive pruning strategy to dynamically remove low contribution Gaussian primitives based on network conditions.
    \item Extensive experiments demonstrate that AirGS reduces PSNR deviation  by more than 20\% when scene changes,  maintains frame-level PSNR consistently above 30, accelerates training by 6$\times$, and reduces per-frame transmission size by nearly 50\% compared to the state-of-the-art 4DGS approaches.

\end{itemize}

The rest of this paper is organized as follows. We present background on 3DGS and existing 4DGS efforts towards FVV in Section \ref{sec:background}. Section \ref{sec:design} details the system design of AirGS. Section \ref{sec:evaluation} presents a comprehensive evaluation of AirGS. Finally, we discuss related work in Section \ref{sec:related} and conclude the paper in Section \ref{sec:conclusion}.

\section{Background}
\label{sec:background}
3D Gaussian Splatting (3DGS)\cite{kerbl20233d} provides high-fidelity 3D reconstruction and novel view synthesis by integrating the strengths of both implicit and explicit rendering. It reconstructs 3D scenes from point clouds generated from Structure-from-Motion (SfM) process \cite{snavely2006photo} by modeling each point as an anisotropic Gaussian ellipsoid. Each ellipsoid or primitive is characterized by a full 3D covariance matrix $\Sigma$, its center spatial position $\mu=(x,y,z)$, its opacity $\alpha$, the color $c$, and its spherical harmonics, $SH$, to reflect the view-dependent effect. The covariance matrix is used to control its geometry, which can be further divided into a scaling matrix $S$ and a rotation matrix $R$. Thus, a Gaussian primitive $i$ can be  defined as $G ^ i = \{\mu ^ i, \alpha ^i, c ^i, SH^i, R^i, S^i\}$. The value of these variables can be obtained through machine learning with a differentiable splatting renderer. The training loss is a combination of mean absolute loss $L_1$ and structural dissimilarity loss $L_{D-SSIM}$: 
\begin{equation}
\label{equ:3DGSloss}
    L_{3DGS} = (1-\lambda)L_1 + \lambda L_{D-SSIM}
\end{equation}
where $\lambda$ is a weighting hyperparameter.

An intuitive approach towards free-viewpoint video construction is to independently train all frames as independent 3DGS models, as illustrated in Fig. \ref{subfig:3dgs}. However, this leads to excessive computational cost. In contrast,
to avoid the complexity and high cost of performing full 3DGS training for every frame, most of existing 4DGS studies construct an anchor space as a reference and then reconstruct the subsequent frames using motion estimation, as illustrated in Fig. \ref{subfig:4dgs}. Specifically, given a sequence of frames, a 3DGS is first trained for the initial frame, forming the canonical Gaussian space $GC_0$ for the entire video. The motion of Gaussian attributes $d_i$ between consecutive frames $(i-1)$ and $i$ is then estimated, allowing the Gaussian representation of any frame $t$ to be derived by the following equation.
\begin{equation}
    GS_t = \sum_{i=0}^n G_t^i = GC_0 + \sum_{i = 1}^td_i
\end{equation}
Streamable 4DGS methods, such as $V^3$ \cite{wang2024v} in Fig. \ref{subfig:airgs}, encode Gaussian primitives into streaming-friendly data formats, such as 2D images. Specifically, all Gaussians for the current frame reconstruction are represented by a set of images, with each image encoding a single attribute, such as opacity, scale, etc. The different attributes of the same Gaussian are stored at the same pixel location across these images.
After streaming all these images to the client, the $GS_t$, which has $n$ Gaussian primitives, can be easily obtained by Equ. \ref{stream_equ} and subsequently employed for local rendering.
\begin{equation}
    GS_t =\sum_{i=0}^n G^i(t) =  \sum_{i = 0}^n\sum_{j = 0}^ma_j(i)
\label{stream_equ}
\end{equation}
where $a_i(j)$ is the value of pixel $i$ in image $j$, and $m$ is the number of attributes, which also corresponds to the number of images transmitted for that frame.

\begin{figure}[!tbp]
  \centering
  \includegraphics[width=\linewidth]{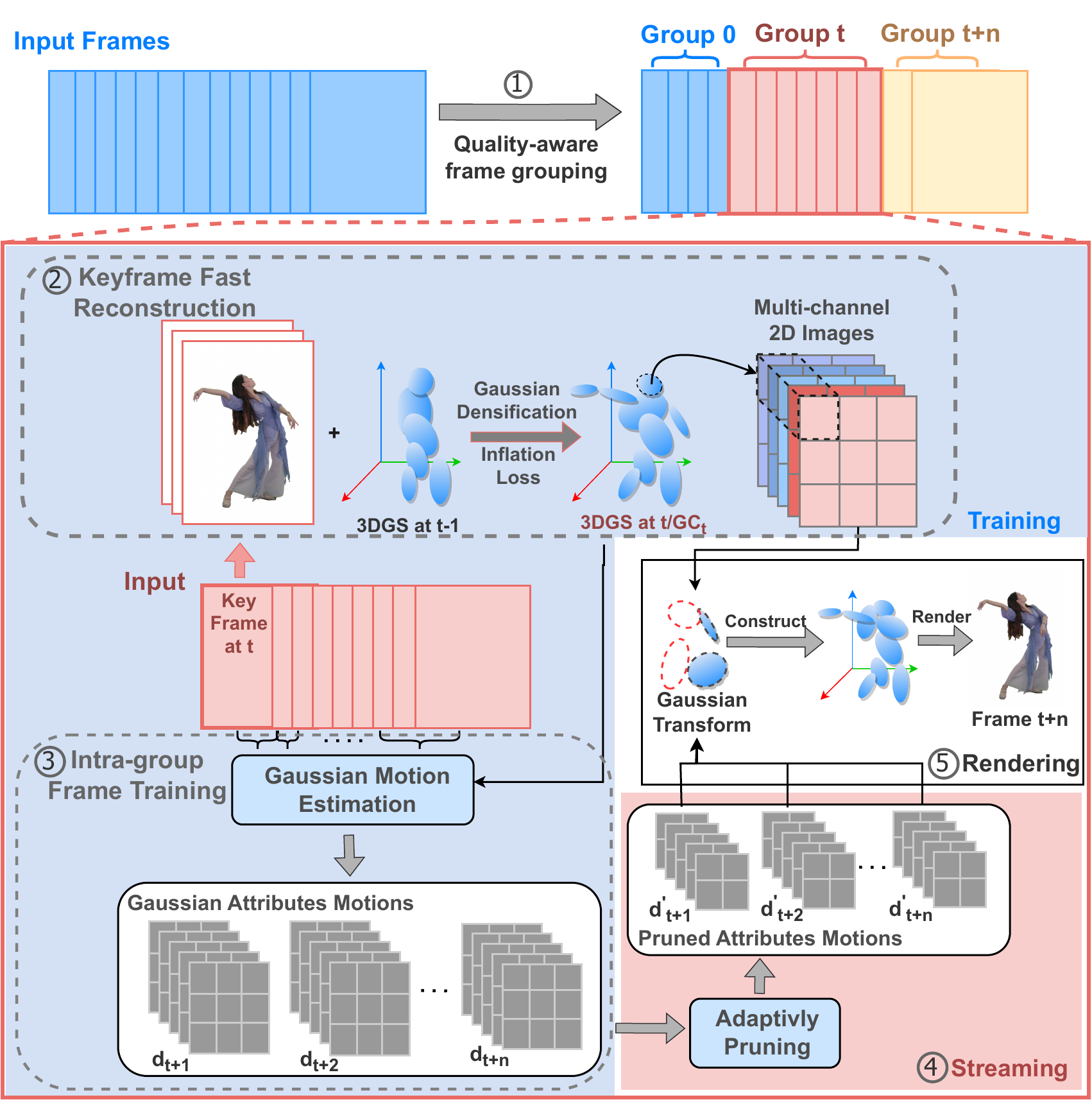}
  \caption{System overview of AirGS. Each frame group is identified by its keyframe index.
  }
  \label{pipeline}
\end{figure}

\section{System Design}
\label{sec:design}
\subsection{System overview}
The system overview of AirGS is shown in Fig. \ref{pipeline}.  
AirGS redesigns both the training and streaming processes of Gaussian primitives. During training, AirGS first groups frames by reconstruction quality, designating any frame below the quality threshold as the keyframe of a new group to capture major scene changes. 
For the keyframe in the first group, AirGS simply follows the traditional 3DGS training strategy. For keyframe reconstruction in subsequent groups, AirGS introduces Gaussian inflation loss combined with Gaussian densification to enable fast, high-quality initialization with little  memory overhead.
For frame training within a group, AirGS builds on existing canonical Gaussian primitive space and leverages  inter-frame similarity to estimate Gaussian motion of the canonical space for reconstruction.
In streaming, rather than transmitting complete Gaussian ellipsoids $GS_t$ at each timestep, AirGS streams the Gaussian differences between adjacent frames, forming a sparse, compression-friendly matrix that significantly reduces transmission cost. To handle fluctuating bandwidth, AirGS applies a usage-frequency-based adaptive pruning strategy to select an appropriate pruning level so that the reconstruction quality can be optimized under the network constraint. 

\subsection{Efficient 4DGS Training for High-Quality Representation}

\begin{figure}[!tbp]
  \centering
  \includegraphics[height=5.3cm, width=0.8\linewidth]{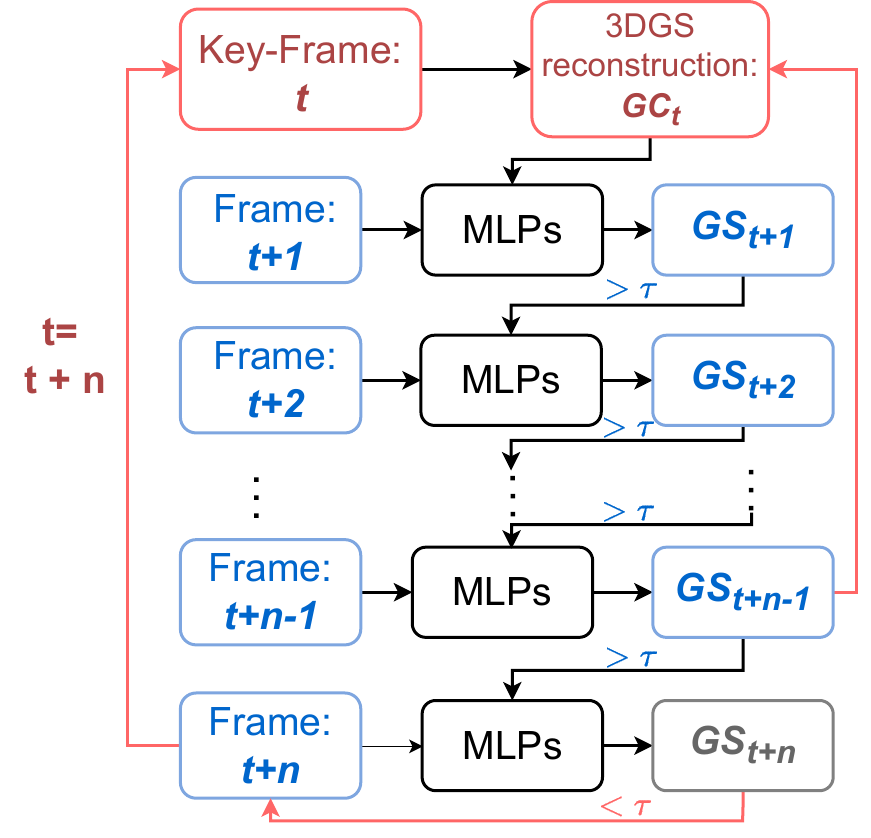}
  \caption{The quality-based frame grouping from group $t$ to the following group $t + n$ based on the key-frame $t$ and quality threshold $\tau$.
  }
  \label{grouping}
\end{figure}

\textbf{The quality-based frame grouping.} 
As discussed in the background, 
once the canonical space is established, the total number of Gaussian primitives remains fixed in existing 3DGS-based solutions for dynamic scene rendering. Consequently, frames containing regions with large gradients during training cannot be fitted by generating new Gaussians, leading to a significant quality decrease.
Therefore, as illustrated in Fig.~\ref{grouping}, we first propose a quality-based frame grouping strategy to iteratively partition entire video sequences into finer-grained groups, 
each with its own canonical space, to enable high-quality FVV construction over long sequence. Unlike prior methods, AirGS first initializes the canonical Gaussian space using the first frame, and subsequent frames are reconstructed by transforming this canonical space. It then employs a quality threshold $\tau$, e.g., value adopted in prior work\cite{lin2024gaussian}\cite{jain1989fundamentals}, to monitor the quality discrepancy between the rendered frame and its ground truth frame during the initial reconstruction. When the reconstruction quality falls below this threshold, indicating that the fixed canonical Gaussians can no longer adequately model the current frame, the frame is designated as another new keyframe and used to initiate a new canonical space for the following frames.

\textbf{Frame training within a group.} 
In 3DGS, obtaining the initial point cloud with SfM method for Gaussian reconstruction of each frame often incurs substantial computational overhead.  Benefited from the temporal continuity brought by our grouping strategy,  AirGS instead choose to reconstruct new frames within a group by fine-tuning the results of the preceding frame, with reconstruction cost significantly reduced. 
Specifically, for two consecutive frames at timestep $t$ and $t{+}1$ within the same group, the optimized reconstruction $GS_t$ is used to initialize $GS_{t+1}$. 
Leveraging the shared canonical space, AirGS predicts inter-frame variations using six MLPs, one for each Gaussian attribute ($\mu$, $\alpha$, $c$, $SH$, $R$, $S$), eliminating full retraining for each frame, ensuring spatial consistency, and accelerating training.
Given the $GS_t$ for frame $t$, the motion $d_{t + 1}$ can be obtained by:
\begin{equation}
    d_{t + 1} = \{MLP_1(GS_t), MLP_2(GS_t), ..., MLP_6(GS_t)\}
\end{equation}
Then, as shown in Fig.~\ref{grouping}, by applying the estimated motion to the $GS_t$, we can efficiently get the Gaussian reconstruction for frame $t + 1$.
\begin{equation}
    GS_{t + 1} = GS_t + d_{t + 1}
\end{equation}
For the training of the MLP, AirGS further introduces a temporal loss $L_{temp}$ as follows to enhance the temporal coherence between adjacent frames. 
\begin{equation}
    L_{temp} = \sum_{i = 1}^6 || MLP_i(GS_t)||_1 
\label{group_temp}
\end{equation}
The loss is designed to prevent disruptions in inter-frame similarity that could lead to perceptual discontinuities. 
Then, combined with the conventional 3DGS loss $L_{3DGS}$ in Equ. (\ref{equ:3DGSloss}), we can get the total loss function for the reconstruction of frames within the same group:
\begin{equation}
    L = L_{3DGS} + \lambda_t L_{temp},
\end{equation}
where $\lambda_t$ is the weight of our temporal loss.

\begin{figure}[!tbp]
  \centering
  \includegraphics[width=\linewidth]{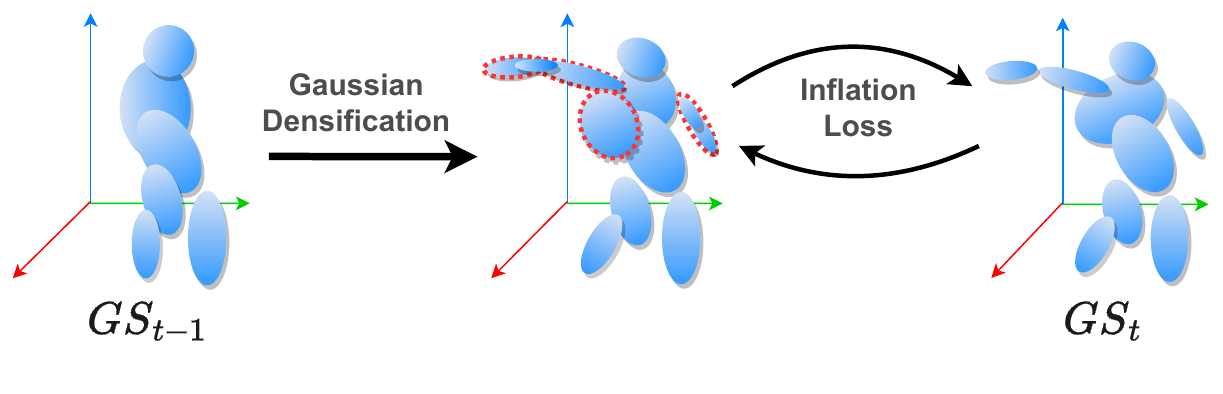}
  \caption{Training process of keyframe $t$. The 3DGS for frame $t-1$ ($GS_{t-1}$) initializes the model. Gaussian densification adds more primitives in high-gradient regions (e.g., red box) for better fitting. An iterative optimization with inflation loss prunes redundancies, yielding the final reconstruction $GS_t$, which serves as the canonical space $GC_t$.
  }
  \label{kf_train}

\end{figure}

\textbf{Keyframe fast reconstruction.}
Keyframe candidates often contain newly emerged objects or large motions, which indicates that the previous 3DGS canonical space can no longer support high-quality reconstruction of the current frame through transformations based on the fixed number of Gaussian primitives alone. Consequently, a new 3DGS canonical space must be re-initialized for accurate reconstruction. However, in existing methods \cite{wang2024v}\cite{wang2023neus2}, these spaces can only be obtained through a full reconstruction pipeline, which incurs substantial computational and time overhead.

Fortunately, we observe that newly emerged objects or large motions typically occupy only a small portion of the entire scene, around 20$\%$ based on our measurements among the HiFi4G dataset\cite{jiang2024hifi4g}, while the majority of the frame, such as the background, remains highly consistent with the previous frame.
Therefore, only a few additional Gaussians are required to capture regions inadequately represented by the previous canonical space. 

Therefore, based on this insight, we still use the trained 3DGS of the preceding frame as an initialization for generating the keyframe's 3DGS in the new group. To accurately capture areas with new objects or large motions, which often exhibit high gradients during training, we apply Gaussian densification by dynamically cloning or splitting primitives. The increased Gaussian aims to represent the scene more precisely, enabling high-quality reconstruction.

However, indiscriminately increasing the number of Gaussian primitives imposes significant computational and storage overhead, particularly for streaming applications, where transmission cost can rise substantially. 
To prevent excessive growth of Gaussian primitives and maximize the rendering efficiency of each one, AirGS introduces an inflation loss during keyframe construction and discards Gaussians with low opacity throughout the training process. The loss progressively reduces the opacity of redundant Gaussian primitives, leading to their eventual elimination as they become visually insignificant.
This process is also illustrated in Fig. \ref{kf_train}. The inflation loss $L_{inf}$ is defined as:
\begin{equation}
    L_{inf} = max(0, N-U),
\end{equation}
where $N$ is the number of current Gaussian primitives and $U$ denotes a predefined soft constraint on the maximum number of Gaussian primitives. In our work, the value of $U$ is determined based on the number of Gaussian primitives in the last canonical space, which is derived from the last keyframe. This design minimizes the risk of exceeding the predefined image capacity, which would otherwise require extra image allocation and cause abrupt increases in data size.

For the training of keyframes, it is also essential to maintain temporal consistency; however, as the MLPs no longer participate in the reconstruction of the keyframes, the temporal loss of key frames $L^k_{temp}$ is defined as
\begin{equation}
    L^k_{temp} = \sum_i||G^i_t -G^i_{t - 1} ||_1,
\end{equation}
where $i$ is bounded by the value of $U$. The inflation loss only removes newly added redundant Gaussians. Since the Gaussians from frame $t{-}1$ are already well-optimized and just insufficient to capture the new content in frame $t$, they are preserved during the keyframe reconstruction, while only redundant additions are pruned. Therefore, for the first frame of the video, its loss follows the conventional 3DGS loss $L_{3DGS}$ in Equation (\ref{equ:3DGSloss}), while for the other following keyframes training, the total loss is defined to be:
\begin{equation}
    L^k_{total} = L_{3DGS} + \lambda_t L^k_{temp} + \lambda_{inf} L_{inf},
\end{equation}
where $\lambda_{inf}$ is the weight of $L_{inf}$, and $k$ is the keyframe's index.

\begin{figure}[!tbp]
  \centering
  \includegraphics[height=4.0cm, width=0.7\linewidth]{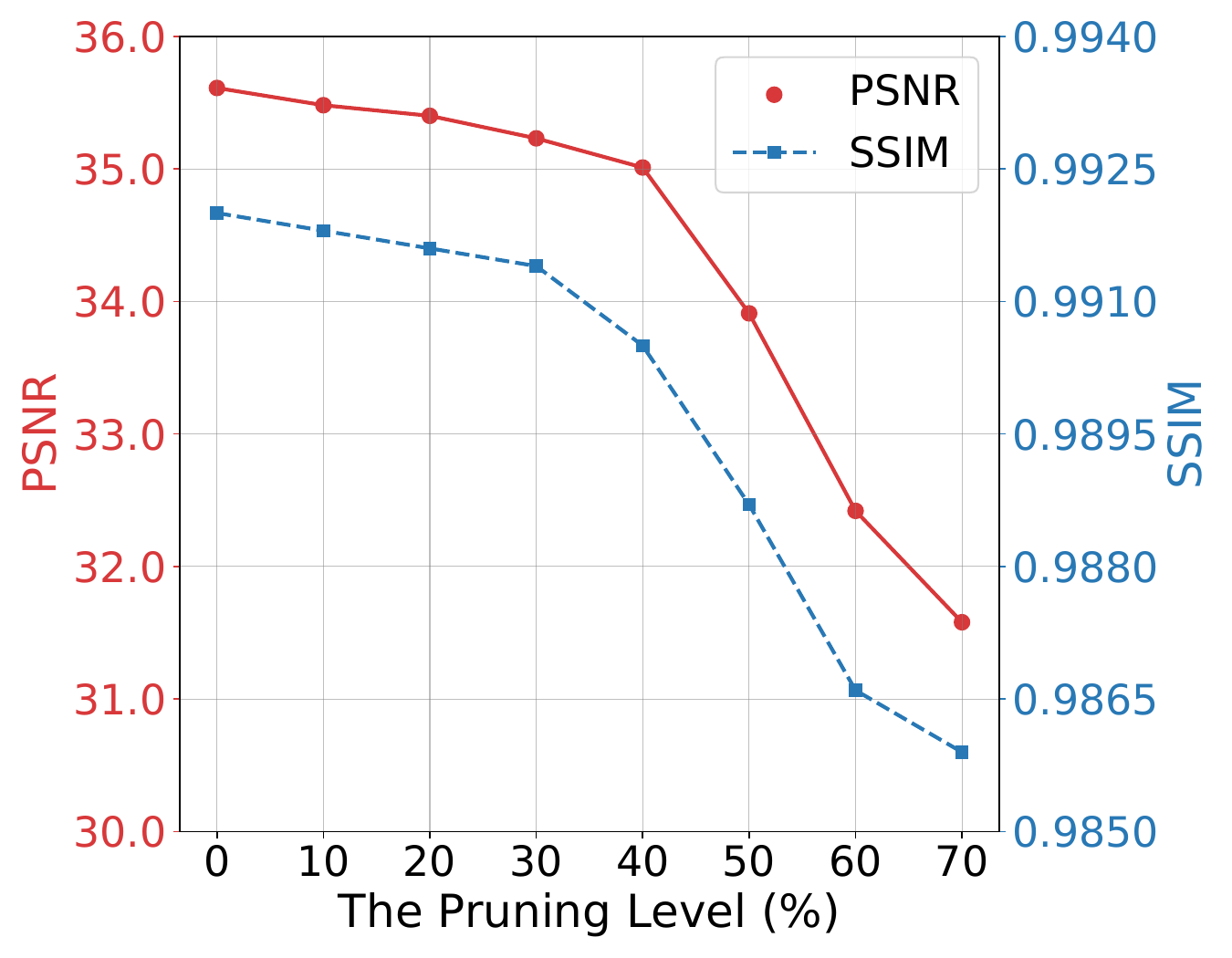}
  \caption{The 3DGS rendering quality with different level of usage-frequency-based Gaussian primitives pruning.}
  \label{prune_quality}
\end{figure}

\begin{algorithm}[t]
  \caption{Pruning level selection for frame $t$}
  \label{Algo}
  \begin{algorithmic}[1]
  \Statex \textbf{Input:} 
  Dense pruning level space $S_t^d$, a list of $(quality, size)$ tuples, each corresponding to a pruning level;
  the "cliff" multiplier, $\beta$;
  the available bandwidth, $B$;
  the target frame rate, $R$
  \Statex \textbf{Output:}
    The pruning level, $p$
  \State $previous \gets S_t^d[1][0] - S_t^d[0][0]$
  \State Initialize an empty array $S$
  \State $S\gets[0]$
  \For{$i$ from 1 to $len(S_t^d) - 1$}
    \State $drop \gets S_t^d[i][0] - S_t^d[i - 1][0]$
    \If{$drop/previous > \beta$}
        \State $p \gets i$
        \State break
    \EndIf
    \State $S$.append($i$)
    \State $previous \gets drop$
  \EndFor
  \State $C\gets B/R$
  \State $low \gets 0$
  \State $high\gets len(S) - 1$
  \While{$low \leq high$}
    \State $mid \gets floor((low + high)/2)$
    \State $tmp \gets S[mid]$
    \If{$S_t^d[tmp][1]\leq C$}
        \State $p\gets tmp$
        \State $high \gets mid - 1$
    \Else
        \State $low\gets mid+1$
    \EndIf
  \EndWhile

  \State \textbf{return} $p$
  \end{algorithmic}
\end{algorithm}

\subsection{Adaptive Pruning for 4DGS Streaming} 
After finishing the training process in remote servers, the data required for rendering, i.e., the attributes of each frame's 3DGS, is transmitted to the client side. In AirGS, two types of data are retained on the server for each video: the multi-channel 2D images for keyframes, where each image stores one Gaussian attribute and each pixel corresponds to one primitive, enabling fast, low-overhead decoding by naturally leveraging hardware codecs for efficient streaming and decoding; the difference tensor $D_t= \sum_{i=k}^td_i$ represents the variation between frame $t$ and its corresponding group's canonical space $GC_k$. This design enables fast reconstruction of any frame within a group, thereby supporting efficient frame skipping. 

To further reduce bandwidth usage during sequential playback, we adopt a differential transmission strategy, wherein only $d_t$ is transmitted for frame $t$ if frame $t-1$ was the last viewed frame. This approach leverages inter-frame similarity, as $d_t$ tends to be sparse matrices, thus reducing both storage and transmission cost. However, under fluctuating network bandwidth, this strategy alone remains insufficient.
As more keyframes are accumulated, the number of Gaussian primitives in the canonical space continuously increases, leading to a progressively larger amount of data $d_t$ that must be transmitted.

To address this, AirGS further adopts an adaptive pruning strategy based on the Gaussian primitive's usage frequency ($p(\cdot)$) to compress the transmission data, which is $d'_t = p(d_t)$.
The goal of the adaptive pruning is to figure out a Gaussian primitives pruning level that maintains streaming smoothness while minimizing degradation in rendering quality under the fluctuation network bandwidth. To prevent cumulative quality degradation caused by pruning-induced variation loss, we maintain a record $D_t' = \sum_t d_t'$ on the server,  which tracks the total amount of data actually transmitted up to frame $t$ from the keyframe. Thus, the data to be sent for frame $t$ within the group $k$ should be:
\begin{equation}
    d'_t = p(D_t - D_{t-1}')
\end{equation}
and the reconstruction follows the following equation:
\begin{equation}
    GS_t = GC_K + \sum_{i=0}^td_t'=GC_k + \sum_{i=0}^t p(D_t-D'_{t-1})
\end{equation}

\textbf{Pruning level selection problem formulation.} Let $n$ be the total number of frames, and $S$ the set of available pruning levels. For a pruning level $j \in S$, let $s_{tj}$ and $q_{tj}$ denote the pruned data size and reconstruction quality of frame $t$, respectively. Given the bandwidth constraint $B_t$ for frame $t$ and a target transmission rate $R$. The pruning level selections for a group of frames can be formulated as follows: 
\begin{equation}
\begin{aligned}
    maximize&\sum^n_{t=1}\sum_{j\in S}q_{tj}x_{tj} \\
    subject\ to&\sum_{j\in S}s_{tj}x_{tj} \leq B_t/R \\
    & \sum_{j\in S}x_{tj} = 1, \forall t  \\
    & x_{tj}\in\{0, 1\}, j\in S \\
    & t \in \{0, 1, \dots, n\}
\end{aligned}
\label{opti_equ}
\end{equation}
where $x_{tj}$ indicates whether pruning level $j$ is selected for streaming the frame $t$. The first constraint limits the data size to be transmitted for frame $t$ under $B_t/R$ to achieve the target frame rate $R$. The second ensures only one pruning level per frame is selected. The last two define the selection variable and total frame count.

\begin{figure*}[!tbp]
  \centering
  \includegraphics[width=\linewidth]{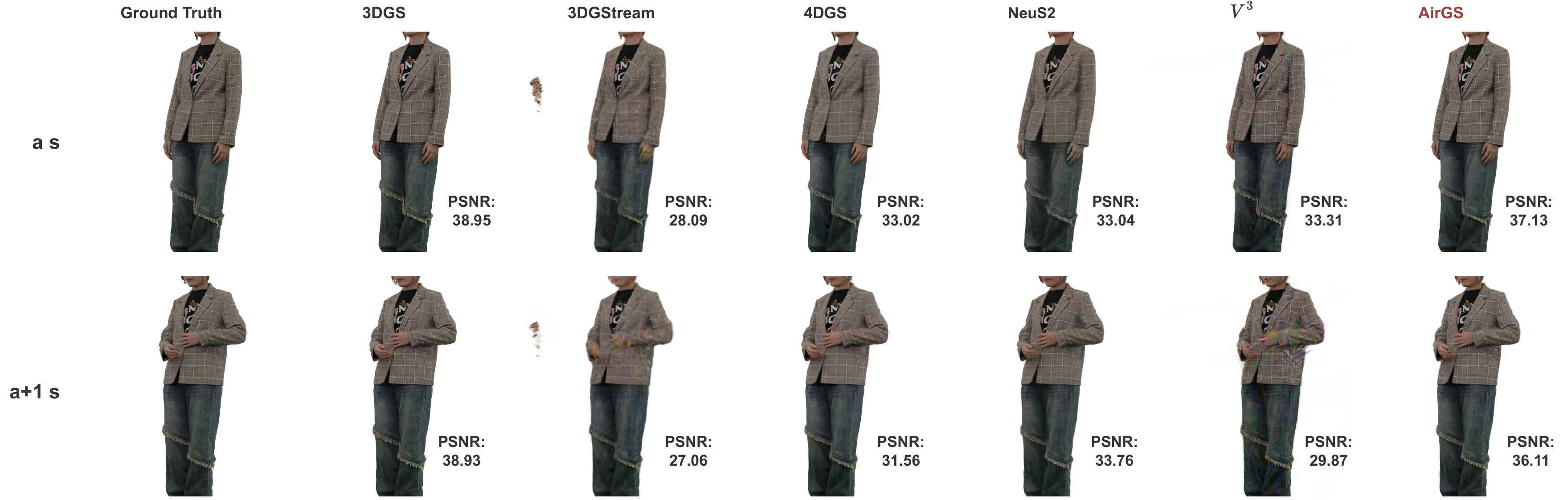}
  \caption{Rendering results comparison of the dynamic scene against recent SOTA methods, showing the reconstruction at time step a and a+1.}
  \label{visualization_d}
\end{figure*}

\begin{table*}[!tbp]
\centering
\caption{Comparison with baselines on dynamic scenes reconstruction. Cells highlighted in red indicate the \colorbox{pink!50}{best} results, while those in yellow denote the \colorbox{yellow!40}{second-best}.}
\begin{tabular}{lccccc}
\toprule
\textbf{Method} & \textbf{PSNR} $\uparrow$ & \textbf{SSIM} $\uparrow$ & \textbf{Size (MB)} $\downarrow$ & \textbf{Training Time (Min)} $\downarrow$ & \textbf{Rendering Smoothness (FPS)} $\uparrow$\\
\midrule
NeuS2 \cite{wang2023neus2}        & 29.82 & 0.961 & 24.5   & 3.7  & 3\\
Naive 3DGS \cite{kerbl20233d}     & \cellcolor{pink!50}33.94 & \cellcolor{pink!50}0.987 & 18.3   & 9.8  & 200\\
4DGS \cite{wu20244d}              & 29.95 & 0.971 & 12.0   & 9    & 60\\
3DGStream \cite{sun20243dgstream} & 25.33 & 0.944 & 7.8    & \cellcolor{pink!40}0.75 & 215\\
$V^3$ \cite{wang2024v}            & 27.69 & 0.949 & \cellcolor{yellow!40}2.0    & 1.5  & \cellcolor{yellow!40}387\\
\rowcolor{gray!10}
\textbf{AirGS} & \cellcolor{yellow!40}32.29 & \cellcolor{pink!50}0.987 & \cellcolor{pink!50}1.1 & \cellcolor{yellow!40}1.4 & \cellcolor{pink!50}400\\
\bottomrule
\label{quantive_table_all}
\end{tabular}
\end{table*}

\textbf{Algorithm design} To efficiently determine an appropriate pruning level, AirGS adopts a stepwise fallback strategy, which selects the lowest pruning level whose post-pruning size cost does not exceed the bandwidth requirement. The next step is to construct the pruning level space $S$. As shown in the Fig. \ref{prune_quality}, pruning generally correlates with quality degradation, with a tipping point where further pruning leads to a sharp quality drop.
Therefore, for frame $t$, we first perform fine-grained pruning based on the rendering usage frequency of each Gaussian primitive. The pruning ratios are varied across a wide range, thereby constructing a dense pruning level space $S_t^d$. Due to the high rendering efficiency of 3DGS, we can rapidly evaluate the reconstruction quality at each pruning level. When the quality degradation between two adjacent levels significantly exceeds $\beta$ times that of previous steps, the latter is identified as a quality cliff, and all preceding levels are retained as the final pruning space $S$. Leveraging the monotonic relationship between pruning levels and quality degradation, we employ binary search to efficiently identify the lowest pruning level that satisfies the bandwidth constraint, thereby maximizing the quality of each frame. 
Thus, the detailed algorithm is shown in the Algorithm \ref{Algo}.

\textbf{Complexity Analysis} This algorithm can be divided into two parts, the first part (line 4 to line 11) scans the dense pruning level space to construct a reduced candidate set $S$.
Pruning levels in $S_t^d$ that meet the quality requirement are inserted into $S$ in increasing pruning order, ensuring a monotonically decreasing size sequence. The time complexity for this candidate set construction part is $O(n)$. The second part aims to identify a feasible pruning level that maximizes the reconstruction quality under the constraints of Equ. \ref{opti_equ}. Thanks to the monotonically decreasing size order of $S$ maintained in the first part, a binary search with a time complexity of $O(log$ $n)$ can be efficiently applied to locate a valid pruning level.

\begin{figure}[!tbp]
  \centering
  \includegraphics[height=5.0cm, width=0.9\linewidth]{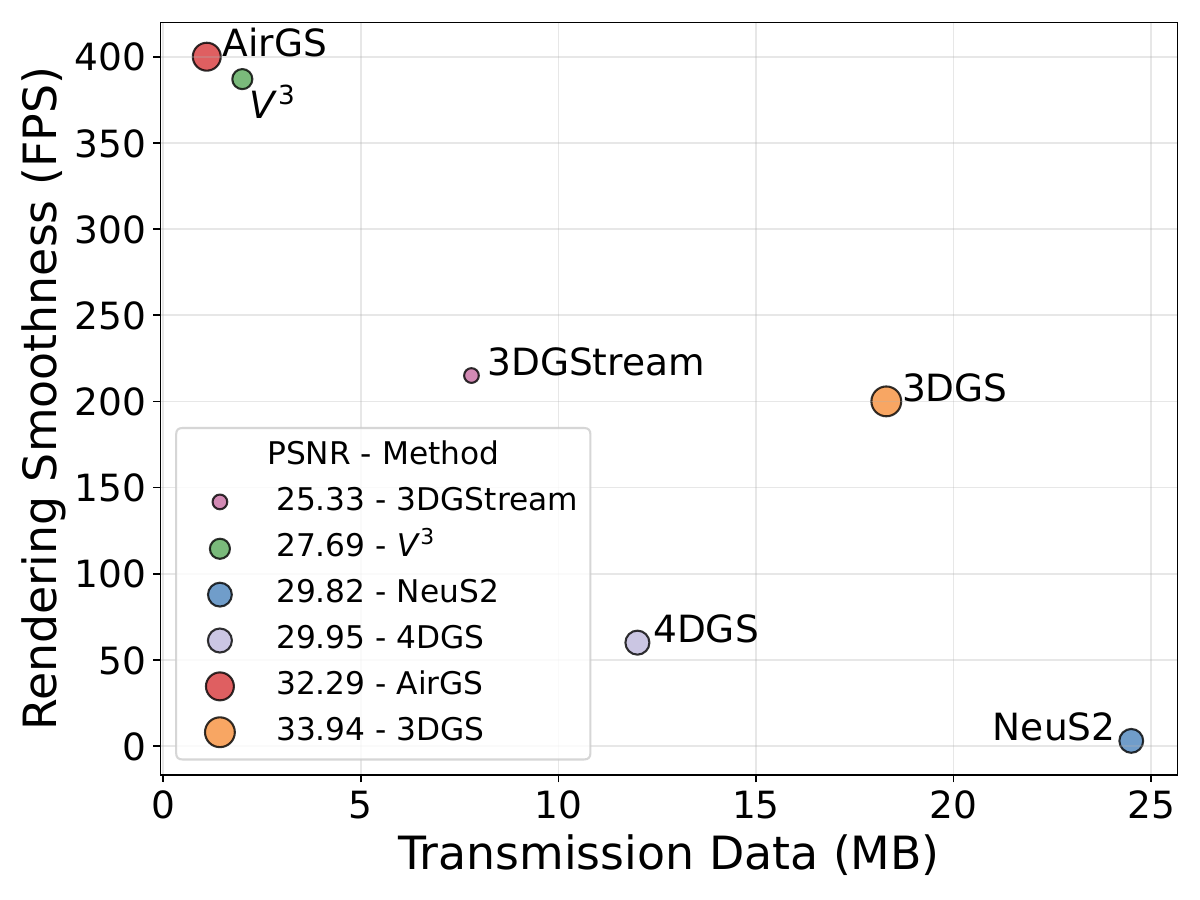}
  \caption{Compared with other baselines, AirGS achieves high-quality and smoother rendering with a small data size required. The point size refers to PSNR.}
  \label{quantitative_metric_figure}
\end{figure}

\section{Evaluation}
\label{sec:evaluation}
\subsection{Evaluation setup}
\textbf{Dataset.} We evaluate AirGS on the HiFi4G dataset \cite{jiang2024hifi4g}, which includes seven dynamic sequences captured at 30 fps from 81 viewpoints with a resolution of 3840*2160. For each sequence, 10 viewpoints are held out for testing, while the remaining views are used for training. To assess streaming performance under practical network conditions, we adopt the driving scenario in LTE traces \cite{raca2018beyond}, which signifies limited bandwidth and significant network fluctuation.

\textbf{Baselines.}
We compare AirGS with several state-of-the-art methods for dynamic scene reconstruction and rendering. For neural radiance field–based approaches, we select NeuS2 \cite{wang2023neus2}, which leverages multi-resolution encoding and progressive training for efficient rendering but may produce blurred details in fine-grained regions.
Among 3DGS-based methods, in addition to the original 3DGS \cite{kerbl20233d}, we also compare AirGS with three other state-of-the-art benchmarks: 3DGStream \cite{sun20243dgstream}, which uses a neural transformation cache to capture inter-frame changes for fast reconstruction; 4DGS \cite{wu20244d}, which fuses Gaussians with 4D neural voxel representations for high-quality rendering; and $V^3$ \cite{wang2024v}, which adopts canonical space and improves streaming efficiency by projecting each frame’s 3DGS into 2D images

\textbf{Metrics.}
We evaluate performance across multiple dimensions that critically impact the user’s viewing experience.
\begin{itemize}
    \item Rendering visual quality: We utilize two widely adopted metrics to evaluate the rendering performance of AirGS.

    \textbf{PSNR:} This metric quantifies the difference between two images by calculating the mean squared error (MSE) of their pixel values. This metric provides a numerical assessment of the differences between the reconstructed frame and the ground truth\cite{hore2010image}.

    \textbf{SSIM:} This metric evaluates the similarity between two images by jointly considering luminance, contrast, and structural information. An SSIM score of 1 indicates that the images are perfectly identical \cite{wang2004image}.
    \item Transmission data size (MB). This metric measures the amount of data required to render a frame, which also reflects the bandwidth consumption per reconstruction.
    \item Training time. This metric reflects the efficiency of training on dynamic scenes using different methods.
    \item Rendering smoothness. Frames per second (FPS) measures rendering fluency and reflects the computational burden of each method, serving as a key indicator of smooth user experience.
    \item Transmission time. This metric measures the time needed to transmit frame data from the server to the client. Higher transmission times can cause playback stalls or interruptions, degrading the viewing experience.
\end{itemize}

\textbf{Hyperparameter Settings.} 
All training and evaluation are conducted on a single NVIDIA GeForce RTX 3090 GPU. For all baselines, the entire video is treated as a single group, and keyframes are selected according to each method’s default strategy. 
For AirGS, the quality threshold for frame grouping is set to 30, commonly considered indicative of high-quality reconstruction \cite{lin2024gaussian}\cite{jain1989fundamentals}\cite{mildenhall2021nerf}. Loss hyper-parameters are set as follows: $\lambda = 0.2$, consistent with prior work \cite{sun20243dgstream}\cite{kerbl20233d}, $\lambda_{t} = 1e-3$, and $\lambda_{inf} = 1e-5$. The dense pruning level space $S_t^d$ ranges from 0\% to 100\% in 10\% increments, $\beta$ is set to 2.

\subsection{Evaluation Results}

\textbf{Overall Performance.}
We first demonstrate visual results rendered by different methods in Fig.~\ref{visualization_d}. These results illustrate that AirGS effectively reconstructs fine-grained details and avoids the blurring, detail loss, and artifacts common when modeling large motions, while others show noticeable degradation. 
We next present the overall comparison results in quantifiable metrics in Table \ref{quantive_table_all} and Fig. \ref{quantitative_metric_figure}, where all reported values represent the average performance.  AirGS ranks second only to naive 3DGS \cite{kerbl20233d} in quality, with a PSNR gap of 1.65 dB and comparable SSIM, and outperforms other baselines by up to 5 dB PSNR and 0.043 SSIM.
Moreover, AirGS significantly outperforms all baselines, including the naive 3DGS, in other two key metrics: transmission size and rendering smoothness. Specifically, AirGS reduces the per-frame transmission data from 18.5 MB in naive 3DGS \cite{kerbl20233d} to just 1.1 MB. Compared to the second-best method $V^3$ \cite{wang2024v}, AirGS archives over 40\% data size reduction from 1.9 MB to 1.1 MB. In terms of rendering smoothness, AirGS delivers an average rendering speed twice that of naive 3DGS \cite{kerbl20233d}. These improvements are enabled by the combination of inflation loss and usage-frequency-based pruning, which effectively control the number of Gaussians per frame and enhance rendering efficiency.
In terms of training time, AirGS takes approximately 1.4 minutes per frame, slightly slower than 3DGStream \cite{sun20243dgstream}, which requires only 0.75 minutes per frame, and ranks second overall, yet it outperforms 3DGStream \cite{sun20243dgstream} in all other metrics. 

Table \ref{keyframe_re} further demonstrates the keyframe training time comparison for the keyframe-based methods. Due to the reliance on implicit neural rendering, NeuS2 \cite{wang2023neus2} incurs much higher training cost and requires 7.6 times the reconstruction time of AirGS, while $V^3$ \cite{wang2024v} takes 4.9 times longer. This efficiency gain primarily stems from AirGS’s leverage of strong inter-frame similarity, which avoids redundant retraining on static regions and substantially reduces overall computation.
However, in Table \ref{quantive_table_all}, AirGS achieves only a modest reduction in average per-frame training time compared with it. This is due to AirGS selecting multiple keyframes to better capture significant variations, whereas $V^3$ \cite{wang2024v} uses a single keyframe.

\begin{figure}[!tbp]
        \centering
        \subfigure[Quality fluctuation in PSNR for different methods.]{\includegraphics[height=3cm,width=0.48\linewidth]{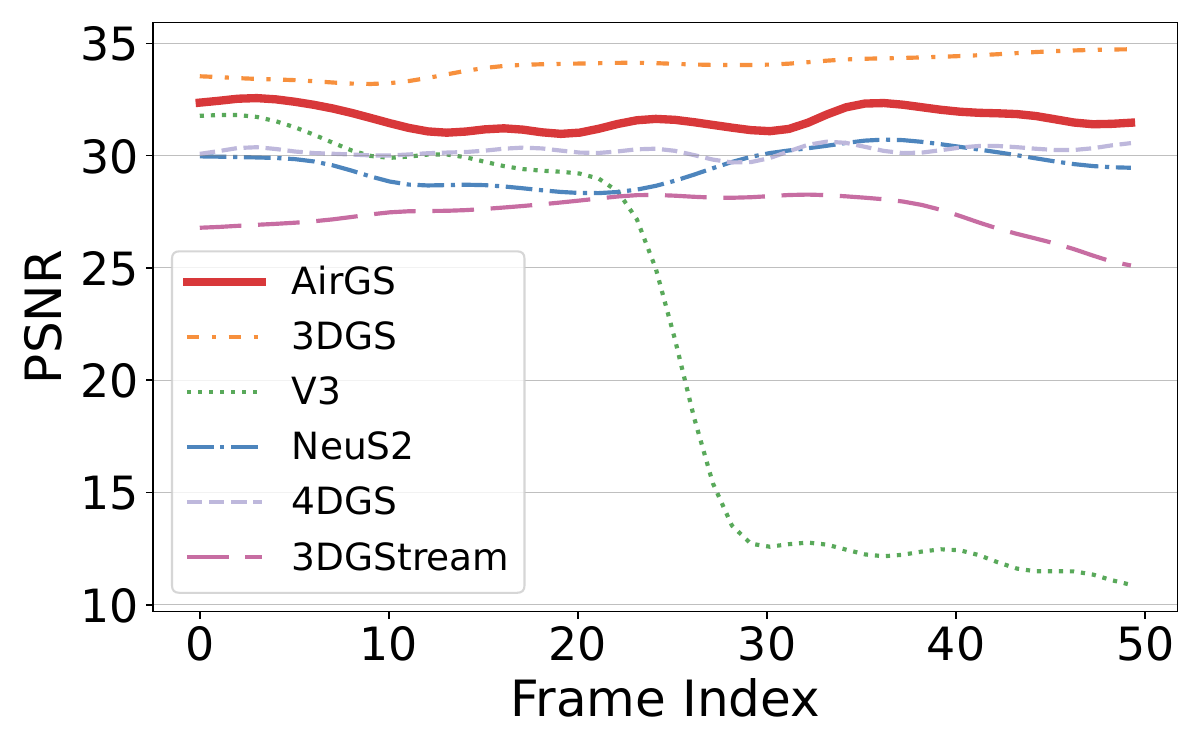} \label{psnr_dynamic}}
        \subfigure[Average quality deviation in PSNR for different methods. The term "GSt." refers to the 3DGStream.]{\includegraphics[height=3cm,width=0.48\linewidth]{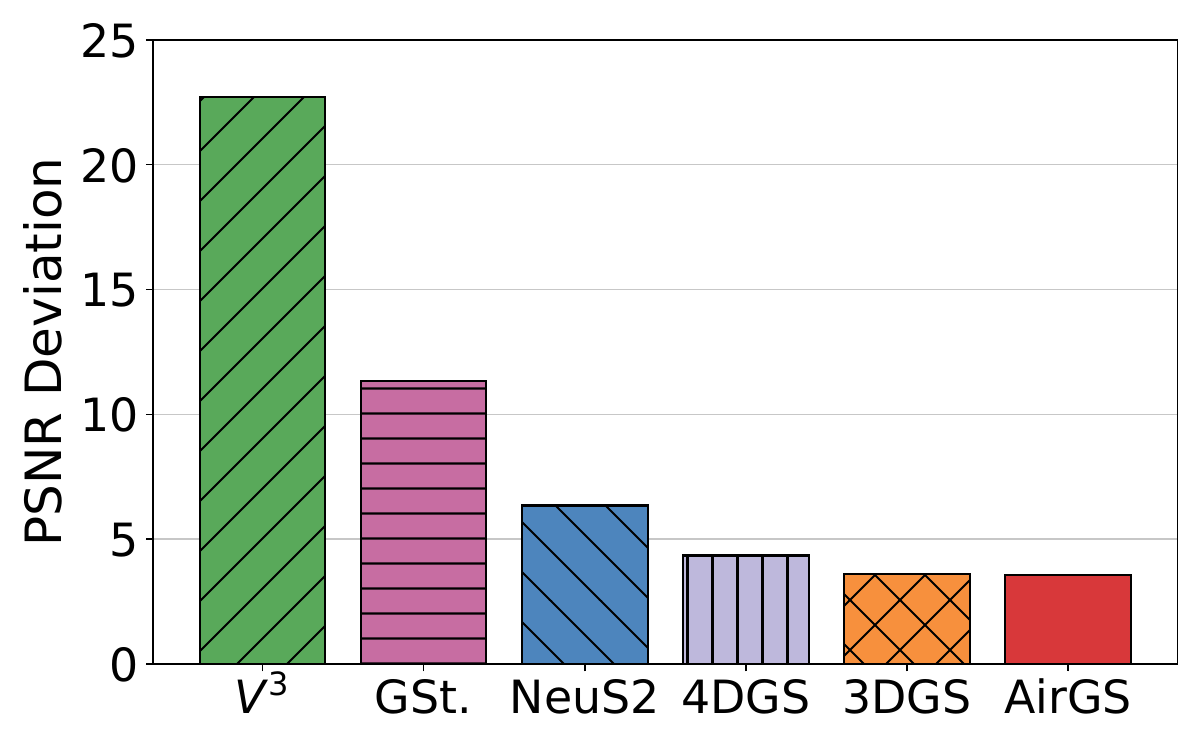}\label{psnr_deviation}}
        \caption{Comparison of quality fluctuation and average quality deviation.}
        \label{quality_sim_scene}
\end{figure}

\textbf{Quality Consistency.}
Fig. \ref{psnr_dynamic} demonstrates the detailed frame-wise PSNR trends across different methods, indicating that AirGS maintains frame-level PSNR above 31, avoids sudden quality drops and accurately captures newly appearing objects and large motions, ensuring high-quality reconstruction. Fig. \ref{psnr_deviation} shows the average maximum per-frame reconstruction deviation across different video sequences. AirGS achieves reconstruction quality comparable to per-frame 3DGS reconstruction, with average PSNR fluctuations below 5 dB, demonstrating that 
by identifying hard-to-fit frames and assigning them new canonical spaces, AirGS’s quality-based keyframe strategy maintains consistently high-fidelity reconstruction over long sequences and prevents the degradation.

\begin{table}[!tbp]
\centering
\caption{Comparison of average keyframe training time}
\begin{tabular}{lc}
\toprule
\textbf{Method} & \textbf{Keyframe training time(min)} $\downarrow$\\
\midrule
NeuS2 \cite{wang2023neus2}        & 4.62 \\
$V^3$ \cite{wang2024v}            & 2.98 \\
\rowcolor{gray!10}
\textbf{AirGS} & \cellcolor{pink!50}0.61 \\
\bottomrule
\label{keyframe_re}
\end{tabular}
\end{table}

\begin{figure}[!tbp]
        \centering
        \subfigure[CDF of frame transmission time among different methods.]{ \includegraphics[height=3cm, width=0.48\linewidth]{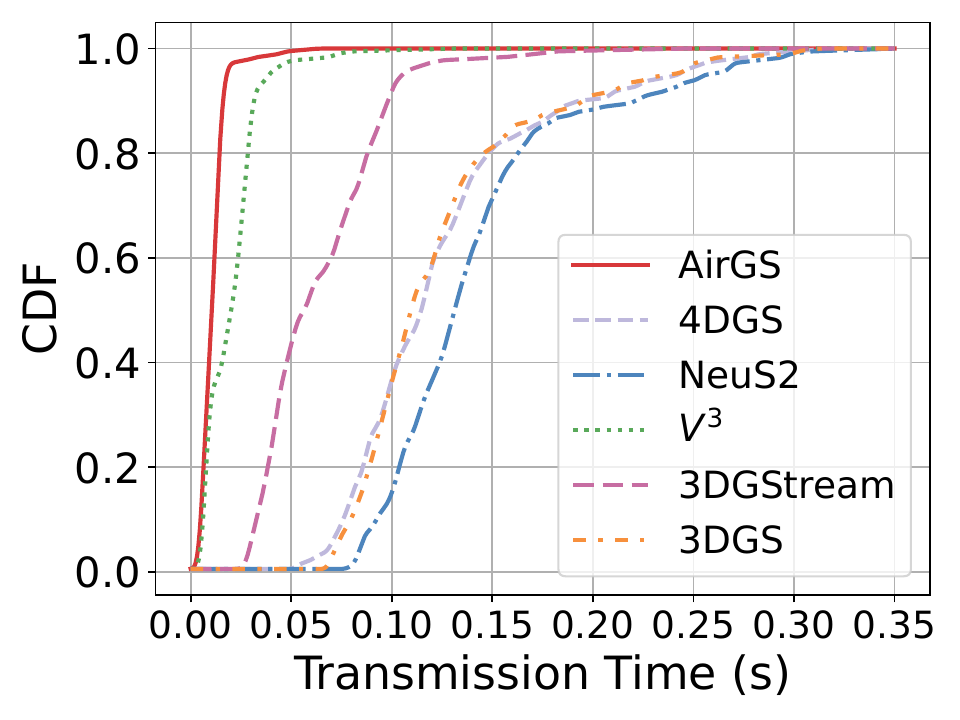} \label{receiving_cdf}}
        \subfigure[Maximum transmission time per frame among different methods. The term "GSt." refers to the 3DGStream.]{\includegraphics[height=3cm,width=0.48\linewidth]{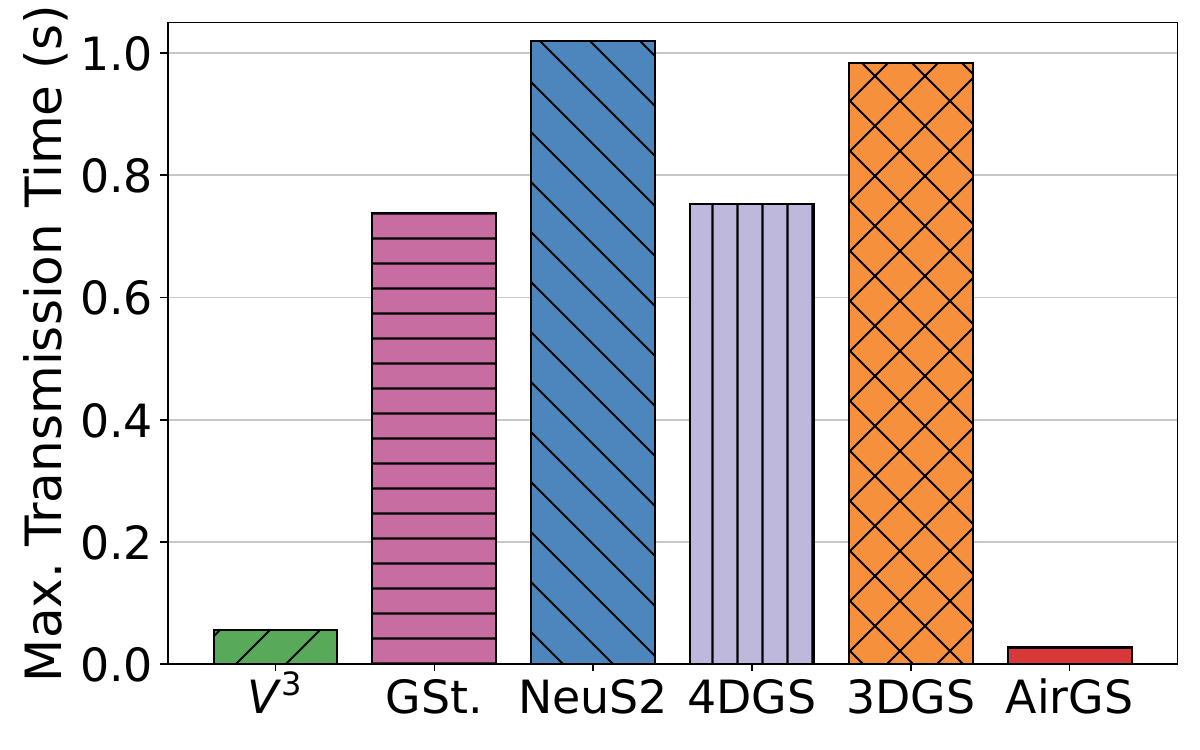}\label{max_tran}}
        \caption{CDF and the maximum value of transmission time of different methods.}
        \label{transmit_perform}
\end{figure}

\textbf{Transmission Performance.}
Fig. \ref{transmit_perform} and Table \ref{streaming} present the data transmission performance of different methods in realistic networks. The cumulative distribution in Fig. \ref{receiving_cdf} shows that, thanks to its adaptive pruning strategy, AirGS can transmit the necessary data for around 95\% of frames within 25 ms, even under poor bandwidth. In contrast, some methods require up to 350 ms per frame. Fig. \ref{max_tran} further demonstrates the maximum transmission time per frame for each method. Compared to all other baselines, AirGS reduces the maximum transmission time by up to 37 times, and even compared to the streaming-optimized $V^3$ \cite{wang2024v}, it requires only half the time.
As shown in Table~\ref{streaming}, AirGS achieves an average per-frame transmission time of just 12ms. Compared to the second fastest method, $V^3$ \cite{wang2024v}, which requires 23 ms per frame, AirGS delivers nearly a twofold speedup. Relative to the transmission time of naive 3DGS \cite{kerbl20233d} and the implicit rendering approach NeuS2 \cite{wang2023neus2}, AirGS requires only 8.8\% and 7.7\% of their time, respectively. The second column of Table~\ref{streaming} reports the average transmission rates (frames/min). AirGS is the only method exceeding 90, substantially higher than the next-best rate of just over 40. Maintaining this high transmission rate is essential for a seamless user viewing experience.

\begin{table}[!tbp]
\centering
\caption{Comparison of Average transmission performance}
\begin{tabular}{lcc}
\toprule
\textbf{Method} & \textbf{Transmission Time (s)} $\downarrow$ & \textbf{Transmission Rate} $\uparrow$\\
\midrule
NeuS2 \cite{wang2023neus2}        & 0.139 & 7.18\\
Naive 3DGS \cite{kerbl20233d}     & 0.122 & 8.18\\
4DGS \cite{wu20244d}              & 0.131 & 8.24\\
3DGStream \cite{sun20243dgstream} & 0.063 & 15.90\\
$V^3$ \cite{wang2024v}            & \cellcolor{yellow!40}0.021 & \cellcolor{yellow!40}48.38\\
\rowcolor{gray!10}
\textbf{AirGS} & \cellcolor{pink!50}0.011 & \cellcolor{pink!50}92.74 \\
\bottomrule
\label{streaming}
\end{tabular}
\end{table}

\begin{figure*}[!tbp]
    \centering
    \subfigure[PSNR = 35.67]{\includegraphics[height=2.8cm,width=0.18\linewidth]{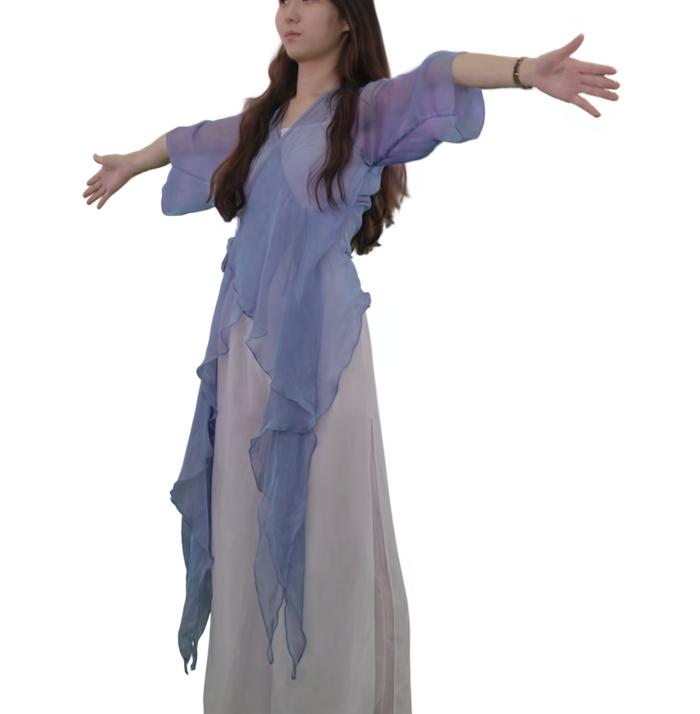}\label{2_no}}
    \subfigure[PSNR = 34.31]{\includegraphics[height=2.8cm,width=0.18\linewidth]{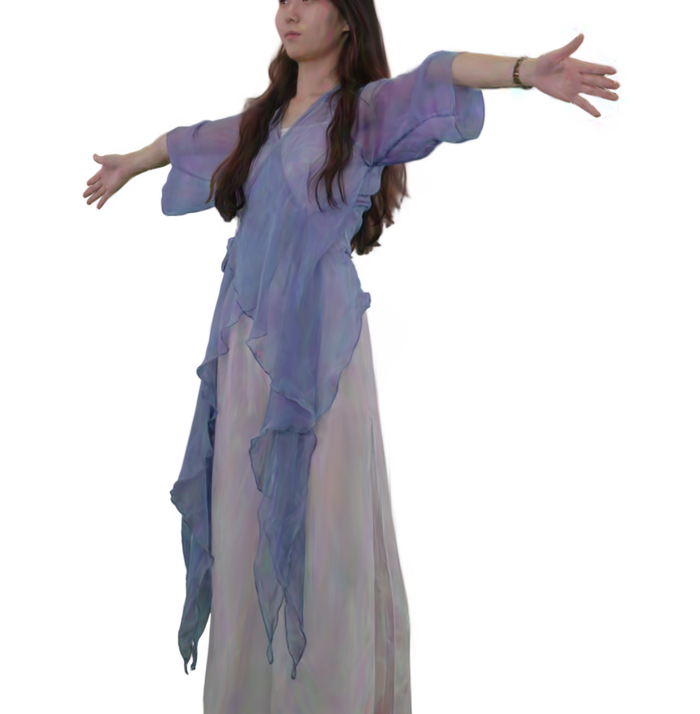}\label{2_max}}
    \subfigure[PSNR = 37.73]{\includegraphics[height=2.8cm,width=0.18\linewidth]{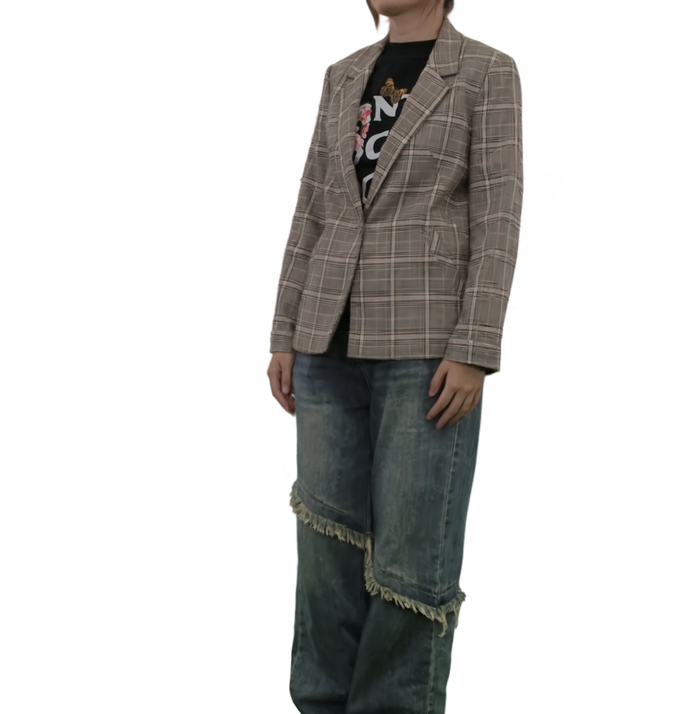}\label{6_n}}
    \subfigure[PSNR = 35.51]{\includegraphics[height=2.8cm,width=0.18\linewidth]{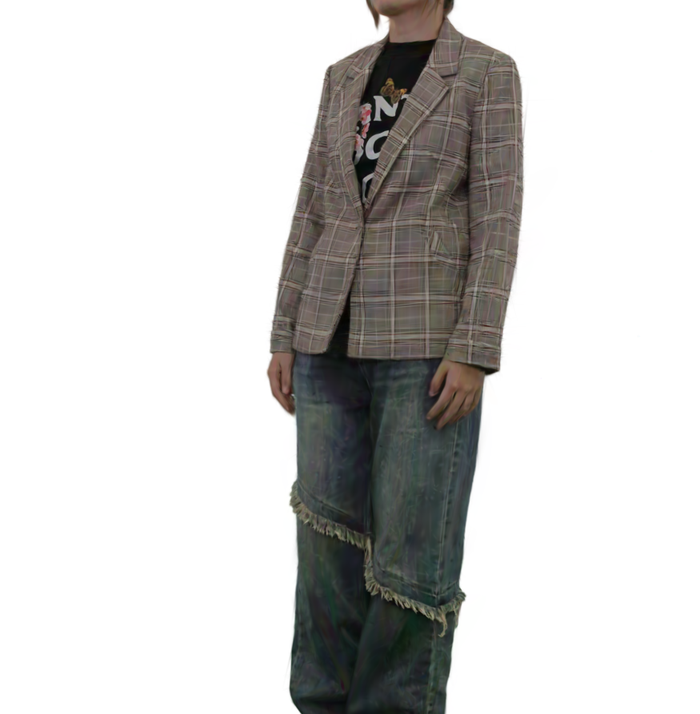}\label{6_max}}
    \subfigure[Average quality and transmission data size comparison]{\includegraphics[height=3cm,width=0.21\linewidth]{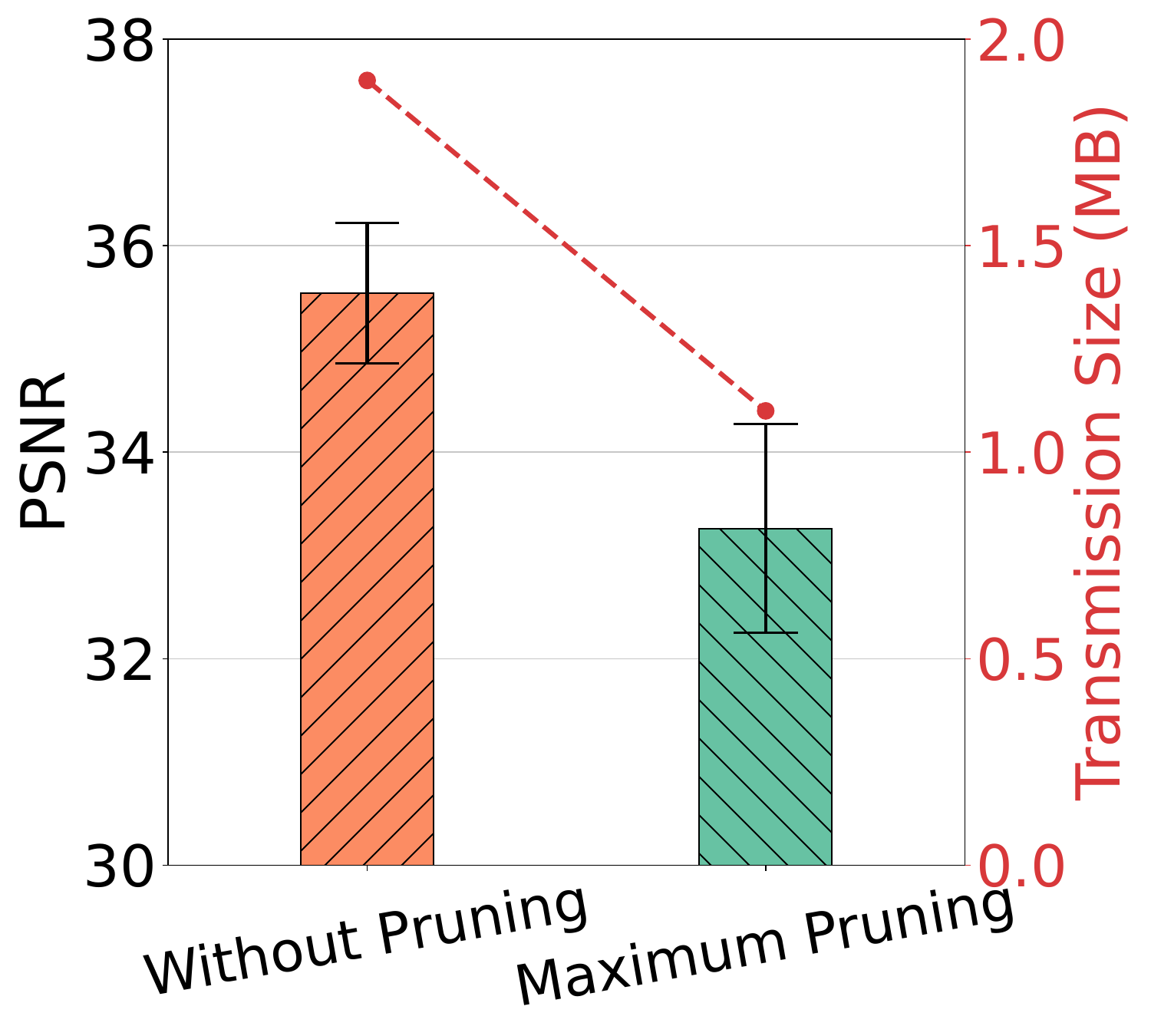}\label{prune_sum}}
    \caption{Impact of pruning on reconstruction quality. The subfigures on the left (a and c) show the rendering results of AirGS without pruning, while those on the right (b and d) illustrate the results after maximum pruning. }
    \label{pruning_level}
\end{figure*}

\begin{figure}[!tbp]
  \centering
  \includegraphics[height=4.0cm, width=0.8\linewidth]{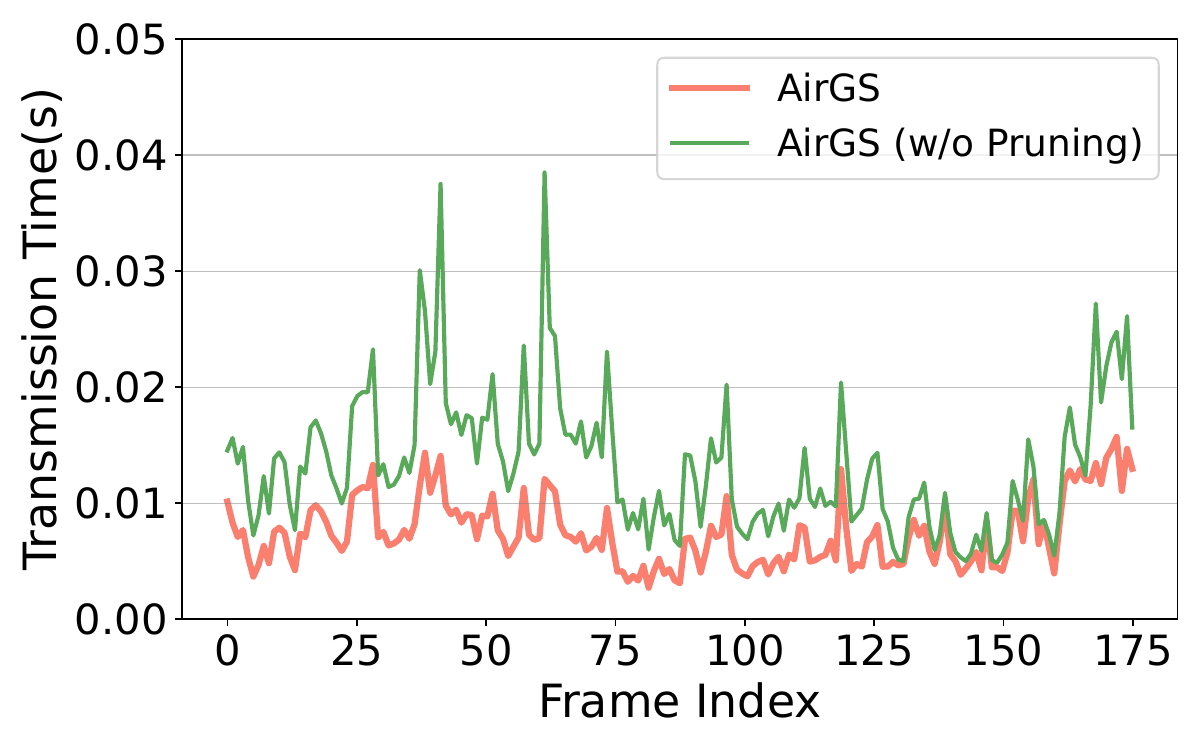}
  \caption{Impact of pruning on transmission time per frame under the LTE network conditions for 175 frames sequences.}
  \label{trans_ab}
\end{figure}

\textbf{Impact of Pruning.} In Fig. \ref{pruning_level}, we demonstrate the impact of pruning on rendering quality. Fig. \ref{2_no} and \ref{6_n} show the rendering results of AirGS without pruning, representing the viewing experience under favorable network conditions. In contrast, Fig. \ref{2_max} and \ref{6_max} depict the visual quality after applying maximum pruning, which serves as the lower bound within the AirGS framework under poor network conditions. Fig. \ref{prune_sum} further illustrates the impact: usage-frequency-based pruning results in an average quality drop of only 2.3, while reducing per-frame transmission size from 1.9 MB to 1.1 MB, nearly a 50\% reduction. 

In Fig. \ref{trans_ab}, we further illustrate the impact of our pruning strategy on transmission time performance. For a 175-frame dynamic scene, pruning reduces the average per-frame transmission time to approximately 0.012s, effectively mitigating large fluctuations caused by network variability. Even in the worst case, transmission remains twice as fast compared to the unpruned setting. These findings demonstrate that our usage-frequency-based pruning strategy effectively preserves high rendering quality, smooth and stable transmission even under adverse network conditions.

\textbf{Overhead Analysis.}
We next evaluate the overhead of AirGS when rendering a frame. The process consists of three phases. The first is streaming, where the data required to render the target frame is transferred. Our measurements show this phase takes approximately 11 ms per frame, accounting for 58.2\% of total processing time. The second phase is decoding, which processes the data into Gaussian differences between the current frame and the target frame, requiring about 25\% of total time.
Finally, the differences are applied to the Gaussian primitives to finish the rendering of the target frame. Overall, AirGS achieves an average end-to-end rendering time of around 18.9 ms per frame, supporting reconstruction at over 50 FPS and delivering a smooth, seamless viewing experience.

\section{Related Work}
\label{sec:related}
\textbf{Novel view synthesis}. Novel view synthesis aims to generate unseen views from multi-view images. Early approaches based on explicit geometric proxies \cite{ma20223d, dalal2024gaussian, penner2017soft, duan20244d} suffered from limited fidelity and high storage costs. NeRF \cite{mildenhall2020nerf}\cite{wang2025nerflex} achieved photorealistic results through implicit rendering but required expensive per-pixel inference, prompting acceleration methods that cache intermediate features in voxels or meshes \cite{hedman2021baking, garbin2021fastnerf, chen2023mobilenerf, wang2022fourier}. Inspired by the hybrid rendering, 3D Gaussian Splatting \cite{kerbl20233d} employs learned Gaussian primitives for high-quality rendering with low computational overhead.

\textbf{Free-Viewpoint Videos of Dynamic Scenes.}
Novel view synthesis for dynamic scenes poses additional challenges due to temporal changes.
Typical methods rely on time-varying primitives \cite{dou2017motion2fusion}\cite{zitnick2004high}\cite{collet2015high}  or interpolation \cite{broxton2020immersive}. NeRF-based approaches model dynamics via canonical templates with time-conditioned deformation \cite{tretschk2021non, fang2022fast, pumarola2021d, park2021nerfies},
scene decomposition
\cite{song2023nerfplayer}, or flow estimation \cite{guo2023forward} \cite{tian2023mononerf}. However, these methods require long training times and suffer from slow rendering. When applied to long video sequences, they often produce artifacts and blur. 
In recent years, 3DGS-based reconstruction emerges as a promising technique for fast and high-quality construction. 
In order to generate real-time FVV, 4DGS \cite{wu20244d}
integrates 3D Gaussian primitives with 4D neural voxel representations to achieve high-resolution, real-time synthesis of dynamic scenes. 3DGStream \cite{sun20243dgstream} uses keyframes, an MLP-based neural transformation cache, and incremental Gaussian splats to warp and update a base scene to the target scene of the specific timestep. 
$V^3$ \cite{wang2024v} encodes per-frame Gaussian attributes into 2D images for full Gaussian primitives streaming, predicts motion deltas with residual entropy and temporal consistency losses based on fixed default frames.

However, these methods struggle to handle scene changes over long sequences because they cannot reliably identify frames with significant variations, causing the default canonical space to become invalid. They are also impractical for real-world network environments, as each frame requires transmitting all Gaussian primitives.
In contrast, AirGS systematically explores training and streaming stages for 4DGS-based FVV, and proposes a multi-keyframe based training framework with a communication-efficient streaming framework.

\section{Conclusion}
\label{sec:conclusion}
In this paper, we present AirGS, a streaming-optimized 4DGS framework that enables high-quality and efficient FVV experience. AirGS integrates a multi-keyframe training framework to preserve reconstruction fidelity in dynamic scenes and an adaptive streaming framework to ensure smooth viewing through dynamic pruning. For training, AirGS employs a quality-driven keyframe selection strategy to initiate new canonical spaces, and leverages inter-frame similarity alongside a set of designed loss functions to facilitate model training.
For streaming, AirGS converts 3DGS data into 2D image representations and adaptively prunes Gaussians to balance reconstruction quality with bandwidth usage under varying network conditions, enabling smooth and responsive rendering. Comprehensive evaluations demonstrate that AirGS consistently outperforms SOTA methods across key metrics, including reconstruction quality, visual consistency, rendering smoothness, training efficiency, and transmission cost.  

\bibliographystyle{IEEEtran}
\bibliography{IEEEexample}

@String{Computer = "{IEEE} Computer" }

@incollection{snavely2006photo,
  title={Photo tourism: exploring photo collections in 3D},
  author={Snavely, Noah and Seitz, Steven M and Szeliski, Richard},
  booktitle={ACM siggraph 2006 papers},
  pages={835--846},
  year={2006}
}

@inproceedings{chen2023mobilenerf,
  title={Mobilenerf: Exploiting the polygon rasterization pipeline for efficient neural field rendering on mobile architectures},
  author={Chen, Zhiqin and Funkhouser, Thomas and Hedman, Peter and Tagliasacchi, Andrea},
  booktitle={Proceedings of the IEEE/CVF Conference on Computer Vision and Pattern Recognition},
  pages={16569--16578},
  year={2023}
}

@inproceedings{garbin2021fastnerf,
  title={Fastnerf: High-fidelity neural rendering at 200fps},
  author={Garbin, Stephan J and Kowalski, Marek and Johnson, Matthew and Shotton, Jamie and Valentin, Julien},
  booktitle={Proceedings of the IEEE/CVF international conference on computer vision},
  pages={14346--14355},
  year={2021}
}

@article{kerbl20233d,
  title={3D Gaussian Splatting for Real-Time Radiance Field Rendering.},
  author={Kerbl, Bernhard and Kopanas, Georgios and Leimk{\"u}hler, Thomas and Drettakis, George},
  journal={ACM Trans. Graph.},
  volume={42},
  number={4},
  pages={139--1},
  year={2023}
}

@article{collet2015high,
  title={High-quality streamable free-viewpoint video},
  author={Collet, Alvaro and Chuang, Ming and Sweeney, Pat and Gillett, Don and Evseev, Dennis and Calabrese, David and Hoppe, Hugues and Kirk, Adam and Sullivan, Steve},
  journal={ACM Transactions on Graphics (ToG)},
  volume={34},
  number={4},
  pages={1--13},
  year={2015},
  publisher={ACM New York, NY, USA}
}

@article{zitnick2004high,
  title={High-quality video view interpolation using a layered representation},
  author={Zitnick, C Lawrence and Kang, Sing Bing and Uyttendaele, Matthew and Winder, Simon and Szeliski, Richard},
  journal={ACM transactions on graphics (TOG)},
  volume={23},
  number={3},
  pages={600--608},
  year={2004},
  publisher={ACM New York, NY, USA}
}

@inproceedings{pumarola2021d,
  title={D-nerf: Neural radiance fields for dynamic scenes},
  author={Pumarola, Albert and Corona, Enric and Pons-Moll, Gerard and Moreno-Noguer, Francesc},
  booktitle={Proceedings of the IEEE/CVF conference on computer vision and pattern recognition},
  pages={10318--10327},
  year={2021}
}

@inproceedings{fang2022fast,
  title={Fast dynamic radiance fields with time-aware neural voxels},
  author={Fang, Jiemin and Yi, Taoran and Wang, Xinggang and Xie, Lingxi and Zhang, Xiaopeng and Liu, Wenyu and Nie{\ss}ner, Matthias and Tian, Qi},
  booktitle={SIGGRAPH Asia 2022 Conference Papers},
  pages={1--9},
  year={2022}
}

@article{song2023nerfplayer,
  title={Nerfplayer: A streamable dynamic scene representation with decomposed neural radiance fields},
  author={Song, Liangchen and Chen, Anpei and Li, Zhong and Chen, Zhang and Chen, Lele and Yuan, Junsong and Xu, Yi and Geiger, Andreas},
  journal={IEEE Transactions on Visualization and Computer Graphics},
  volume={29},
  number={5},
  pages={2732--2742},
  year={2023},
  publisher={IEEE}
}

@inproceedings{guo2023forward,
  title={Forward flow for novel view synthesis of dynamic scenes},
  author={Guo, Xiang and Sun, Jiadai and Dai, Yuchao and Chen, Guanying and Ye, Xiaoqing and Tan, Xiao and Ding, Errui and Zhang, Yumeng and Wang, Jingdong},
  booktitle={Proceedings of the IEEE/CVF International Conference on Computer Vision},
  pages={16022--16033},
  year={2023}
}

@inproceedings{tian2023mononerf,
  title={Mononerf: Learning a generalizable dynamic radiance field from monocular videos},
  author={Tian, Fengrui and Du, Shaoyi and Duan, Yueqi},
  booktitle={Proceedings of the IEEE/CVF International Conference on Computer Vision},
  pages={17903--17913},
  year={2023}
}

@inproceedings{park2021nerfies,
  title={Nerfies: Deformable neural radiance fields},
  author={Park, Keunhong and Sinha, Utkarsh and Barron, Jonathan T and Bouaziz, Sofien and Goldman, Dan B and Seitz, Steven M and Martin-Brualla, Ricardo},
  booktitle={Proceedings of the IEEE/CVF international conference on computer vision},
  pages={5865--5874},
  year={2021}
}

@inproceedings{wu20244d,
  title={4d gaussian splatting for real-time dynamic scene rendering},
  author={Wu, Guanjun and Yi, Taoran and Fang, Jiemin and Xie, Lingxi and Zhang, Xiaopeng and Wei, Wei and Liu, Wenyu and Tian, Qi and Wang, Xinggang},
  booktitle={Proceedings of the IEEE/CVF conference on computer vision and pattern recognition},
  pages={20310--20320},
  year={2024}
}

@article{dou2017motion2fusion,
  title={Motion2fusion: Real-time volumetric performance capture},
  author={Dou, Mingsong and Davidson, Philip and Fanello, Sean Ryan and Khamis, Sameh and Kowdle, Adarsh and Rhemann, Christoph and Tankovich, Vladimir and Izadi, Shahram},
  journal={ACM Transactions on Graphics (ToG)},
  volume={36},
  number={6},
  pages={1--16},
  year={2017},
  publisher={ACM New York, NY, USA}
}

@article{broxton2020immersive,
  title={Immersive light field video with a layered mesh representation},
  author={Broxton, Michael and Flynn, John and Overbeck, Ryan and Erickson, Daniel and Hedman, Peter and Duvall, Matthew and Dourgarian, Jason and Busch, Jay and Whalen, Matt and Debevec, Paul},
  journal={ACM Transactions on Graphics (TOG)},
  volume={39},
  number={4},
  pages={86--1},
  year={2020},
  publisher={ACM New York, NY, USA}
}

@inproceedings{sun20243dgstream,
  title={3dgstream: On-the-fly training of 3d gaussians for efficient streaming of photo-realistic free-viewpoint videos},
  author={Sun, Jiakai and Jiao, Han and Li, Guangyuan and Zhang, Zhanjie and Zhao, Lei and Xing, Wei},
  booktitle={Proceedings of the IEEE/CVF Conference on Computer Vision and Pattern Recognition},
  pages={20675--20685},
  year={2024}
}

@inproceedings{newcombe2015dynamicfusion,
  title={Dynamicfusion: Reconstruction and tracking of non-rigid scenes in real-time},
  author={Newcombe, Richard A and Fox, Dieter and Seitz, Steven M},
  booktitle={Proceedings of the IEEE conference on computer vision and pattern recognition},
  pages={343--352},
  year={2015}
}

@article{wang2024v,
  title={V\^{} 3: Viewing Volumetric Videos on Mobiles via Streamable 2D Dynamic Gaussians},
  author={Wang, Penghao and Zhang, Zhirui and Wang, Liao and Yao, Kaixin and Xie, Siyuan and Yu, Jingyi and Wu, Minye and Xu, Lan},
  journal={ACM Transactions on Graphics (TOG)},
  volume={43},
  number={6},
  pages={1--13},
  year={2024},
  publisher={ACM New York, NY, USA}
}

@inproceedings{tretschk2021non,
  title={Non-rigid neural radiance fields: Reconstruction and novel view synthesis of a dynamic scene from monocular video},
  author={Tretschk, Edgar and Tewari, Ayush and Golyanik, Vladislav and Zollh{\"o}fer, Michael and Lassner, Christoph and Theobalt, Christian},
  booktitle={Proceedings of the IEEE/CVF International Conference on Computer Vision},
  pages={12959--12970},
  year={2021}
}

@inproceedings{wang2022fourier,
  title={Fourier plenoctrees for dynamic radiance field rendering in real-time},
  author={Wang, Liao and Zhang, Jiakai and Liu, Xinhang and Zhao, Fuqiang and Zhang, Yanshun and Zhang, Yingliang and Wu, Minye and Yu, Jingyi and Xu, Lan},
  booktitle={Proceedings of the IEEE/CVF Conference on Computer Vision and Pattern Recognition},
  pages={13524--13534},
  year={2022}
}

@article{dalal2024gaussian,
  title={Gaussian splatting: 3d reconstruction and novel view synthesis, a review},
  author={Dalal, Anurag and Hagen, Daniel and Robbersmyr, Kjell G and Knausg{\aa}rd, Kristian Muri},
  journal={IEEE Access},
  year={2024},
  publisher={IEEE}
}

@inproceedings{duan20244d,
  title={4d-rotor gaussian splatting: towards efficient novel view synthesis for dynamic scenes},
  author={Duan, Yuanxing and Wei, Fangyin and Dai, Qiyu and He, Yuhang and Chen, Wenzheng and Chen, Baoquan},
  booktitle={ACM SIGGRAPH 2024 Conference Papers},
  pages={1--11},
  year={2024}
}

@book{ma20223d,
  title={3D Deep Learning with Python: Design and develop your computer vision model with 3D data using PyTorch3D and more},
  author={Ma, Xudong and Hegde, Vishakh and Yolyan, Lilit},
  year={2022},
  publisher={Packt Publishing Ltd}
}

@inproceedings{mildenhall2020nerf,
 title={NeRF: Representing Scenes as Neural Radiance Fields for View Synthesis},
 author={Ben Mildenhall and Pratul P. Srinivasan and Matthew Tancik and Jonathan T. Barron and Ravi Ramamoorthi and Ren Ng},
 year={2020},
 booktitle={ECCV},
}

@inproceedings{li2021neural,
  title={Neural scene flow fields for space-time view synthesis of dynamic scenes},
  author={Li, Zhengqi and Niklaus, Simon and Snavely, Noah and Wang, Oliver},
  booktitle={Proceedings of the IEEE/CVF Conference on Computer Vision and Pattern Recognition},
  pages={6498--6508},
  year={2021}
}

@inproceedings{wang2025nerflex,
  title={NeRFlex: Resource-aware Real-time High-quality Rendering of Complex Scenes on Mobile Devices},
  author={Wang, Zhe and Zhu, Yifei},
  booktitle={Proc. IEEE ICDCS},
  year={2025}
}

@inproceedings{hore2010image,
  title={Image quality metrics: PSNR vs. SSIM},
  author={Hore, Alain and Ziou, Djemel},
  booktitle={2010 20th international conference on pattern recognition},
  pages={2366--2369},
  year={2010},
  organization={IEEE}
}

@inproceedings{jiang2024hifi4g,
  title={Hifi4g: High-fidelity human performance rendering via compact gaussian splatting},
  author={Jiang, Yuheng and Shen, Zhehao and Wang, Penghao and Su, Zhuo and Hong, Yu and Zhang, Yingliang and Yu, Jingyi and Xu, Lan},
  booktitle={Proceedings of the IEEE/CVF conference on computer vision and pattern recognition},
  pages={19734--19745},
  year={2024}
}

@inproceedings{raca2018beyond,
  title={Beyond throughput: A 4G LTE dataset with channel and context metrics},
  author={Raca, Darijo and Quinlan, Jason J and Zahran, Ahmed H and Sreenan, Cormac J},
  booktitle={Proceedings of the 9th ACM multimedia systems conference},
  pages={460--465},
  year={2018}
}

@inproceedings{wang2023neus2,
  title={Neus2: Fast learning of neural implicit surfaces for multi-view reconstruction},
  author={Wang, Yiming and Han, Qin and Habermann, Marc and Daniilidis, Kostas and Theobalt, Christian and Liu, Lingjie},
  booktitle={Proceedings of the IEEE/CVF International Conference on Computer Vision},
  pages={3295--3306},
  year={2023}
}

@article{mildenhall2021nerf,
  title={Nerf: Representing scenes as neural radiance fields for view synthesis},
  author={Mildenhall, Ben and Srinivasan, Pratul P and Tancik, Matthew and Barron, Jonathan T and Ramamoorthi, Ravi and Ng, Ren},
  journal={Communications of the ACM},
  volume={65},
  number={1},
  pages={99--106},
  year={2021},
  publisher={ACM New York, NY, USA}
}

@book{jain1989fundamentals,
  title={Fundamentals of digital image processing},
  author={Jain, Anil K},
  year={1989},
  publisher={Prentice-Hall, Inc.}
}

@inproceedings{lin2024gaussian,
  title={Gaussian-flow: 4d reconstruction with dynamic 3d gaussian particle},
  author={Lin, Youtian and Dai, Zuozhuo and Zhu, Siyu and Yao, Yao},
  booktitle={Proceedings of the IEEE/CVF Conference on Computer Vision and Pattern Recognition},
  pages={21136--21145},
  year={2024}
}

@article{yang2023real,
  title={Real-time photorealistic dynamic scene representation and rendering with 4d gaussian splatting},
  author={Yang, Zeyu and Yang, Hongye and Pan, Zijie and Zhang, Li},
  journal={arXiv preprint arXiv:2310.10642},
  year={2023}
}

@article{penner2017soft,
  title={Soft 3d reconstruction for view synthesis},
  author={Penner, Eric and Zhang, Li},
  journal={ACM Transactions on Graphics (TOG)},
  volume={36},
  number={6},
  pages={1--11},
  year={2017},
  publisher={ACM New York, NY, USA}
}

@article{wang2004image,
  title={Image quality assessment: from error visibility to structural similarity},
  author={Wang, Zhou and Bovik, Alan C and Sheikh, Hamid R and Simoncelli, Eero P},
  journal={IEEE transactions on image processing},
  volume={13},
  number={4},
  pages={600--612},
  year={2004},
  publisher={IEEE}
}

@inproceedings{hedman2021baking,
  title={Baking neural radiance fields for real-time view synthesis},
  author={Hedman, Peter and Srinivasan, Pratul P and Mildenhall, Ben and Barron, Jonathan T and Debevec, Paul},
  booktitle={Proceedings of the IEEE/CVF International Conference on Computer Vision},
  pages={5875--5884},
  year={2021}
}

\end{document}